\documentclass{aa}

\usepackage{graphicx}

\usepackage[varg]{txfonts}

\usepackage{natbib}
\bibpunct{(}{)}{;}{a}{}{,} % to follow the A&A style

\begin{document}

   \title{Mutual influence of supernovae and molecular clouds}

%   \subtitle{A numerical view of supernova feedback in turbulent medium}

   \author{Olivier Iffrig\inst{\ref{inst1}}
      \and
      Patrick Hennebelle\inst{\ref{inst1}, \ref{inst2}}}

   \institute{Laboratoire AIM, Paris-Saclay, CEA/IRFU/SAp -- CNRS --
      Universit\'e Paris Diderot, 91191 Gif-sur-Yvette Cedex, France
      \label{inst1}
      \and
      LERMA (UMR CNRS 8112), Ecole Normale Sup\'erieure, 75231 Paris Cedex,
      France
      \label{inst2}
      }

   \date{Received July 8, 2014 ; accepted October 29, 2014}

% \abstract{}{}{}{}{}
% 5 {} token are mandatory

  \abstract
     % context heading (optional), leave it empty if necessary
     {Molecular clouds are known to be turbulent and strongly affected by
     stellar feedback. Moreover, stellar feedback is believed to drive
     turbulence at large scales in galaxies.}
     {We study the role played by supernovae in molecular clouds and the
     influence of the magnetic field on this process.}
     {We performed three-dimensional numerical simulations of supernova
     explosions, in and near turbulent self-gravitating molecular clouds. In
     order to study the influence of the magnetic field, we performed both
     hydrodynamical and magnetohydrodynamical simulations. We also ran a
     series of simple uniform density medium simulations and developed a
     simple analytical model.}
     {We find that the total amount of momentum that is delivered during
     supernova explosions typically varies by a factor of about 2, even when
     the gas density changes by 3 orders of magnitude. However, the amount of
     momentum delivered to the dense gas varies by almost a factor of 10 if
     the supernova explodes within or outside the molecular cloud. The
     magnetic field has little influence on the total amount of momentum
     injected by the supernova explosions but increases the momentum injected
     into the dense gas.}
     {Supernovae that explode inside molecular clouds remove a significant
     fraction of the cloud mass. Supernovae that explode outside have a
     limited influence on the cloud. It is thus essential to know
     sufficiently well the correlation between supernovae and the surrounding
     dense material in order to know whether supernovae can regulate star
     formation effectively.}

   \keywords{
         ISM: clouds
      -- ISM: magnetic fields
      -- ISM: structure
      -- ISM: supernova remnants
      -- Turbulence
      -- Stars: formation
   }

   \maketitle

%________________________________________________________________

\section{Introduction}

   Understanding what regulates star formation within galaxies remains an
   unsolved problem.  While it seems clear that it is a consequence of
   fundamental processes such as gravity, turbulence, and magnetic field, it has
   become clear that stellar feedback plays a major role both in limiting the
   star formation efficiency within actively star forming clouds and in
   triggering large-scale interstellar turbulence. In turn, turbulence together
   with gravity contribute to regulating the generation of star forming
   molecular clouds.

   Among other processes such as ionising radiation
   \citep[e.g.,][]{matzner2002}, stellar outflows, or stellar winds, supernovae
   \citep{maclow+2004} are thought to play a fundamental role because they
   contain a very significant amount of energy and momentum. Many studies that
   have attempted to simulate a galactic disk have included supernova feedback
   both at the scale of a whole galaxy \citep[e.g.,][]{tasker-bryan2006,
   dubois+2008, bournaud+2010, kim+2011, dobbs+2011, tasker2011,
   hopkins+2011, renaud+2013}, and at the kpc scale \citep{slyz+2005,
   deavillez+2005, joung+2006,hill+2012, kim+2011, kim+2013, gent+2013,
   hennebelle-iffrig2014}.  In these models it is typically found that the
   supernova feedback can efficiently trigger large-scale turbulence in the
   interstellar medium (ISM) leading to a velocity dispersion on the order of
   $6$ -- $8\ \mathrm{km\ s^{-1}}$.  It is also found that supernovae can
   prevent runaway collapse, leading to star formation rates that are in
   reasonable agreement with observed values \citep{kim+2011, kim+2013,
   hennebelle-iffrig2014}.

   To model supernovae in galactic environments, different strategies have been
   developed. The most fundamental problem is due to the somewhat coarse
   resolution that is employed in these large-scale studies (going from
   typically a few tens of parsecs to a few parsecs). It is not possible to
   follow the supernova explosions with the resolution required to describe them
   accurately and in particular to compute properly the impact they have on the
   gas. More precisely, during the Sedov phase (see below for further
   description), the hot gas that fills the supernova bubble is adiabatic and
   thermal energy is being converted into kinetic energy as the expansion
   proceeds. When the cooling time of the dense shell that surrounds the bubble
   becomes comparable to the dynamical time, the thermal energy is efficiently
   radiated away and the propagation is driven by the shell momentum which stays
   nearly constant. Knowing precisely this momentum is important in the context
   of interstellar turbulence and its replenishment. While some authors have
   directly deposited thermal energy in few grid cells around the supernova
   location \citep[e.g.,][]{deavillez+2005, joung+2006, hill+2012}, some others
   have directly considered the momentum injected \citep[e.g.,][]{kim+2011,
   kim+2013} in the gas by inserting a velocity field that is radial and
   diverging. While the first approach may be more self-consistent, it leads to
   very small time steps because of the very high temperatures ($10^7$ -- $10^8\
   \mathrm{K}$) reached in the simulations.  Moreover, it is generally assumed
   that the supernova remnants are initially spherical at scales comparable to
   the resolution which is typically a few pc. Whether that is a fair assumption
   may depend on the exact place where the supernova is exploding.

   The way to implant supernova feedback in these simulations is largely based
   on the specific models that have been developed to study the evolution of
   supernova remnants in a uniform medium \citep{Oort51, Sedov59, Chevalier74,
   mckee-ostriker1977, Cioffi88, Ostriker88, blondin+1998}. Three stages are
   commonly recognized for the evolution of supernova remnants. The first phase
   is a free expansion phase, where the initial ejecta simply sweep up the ISM.
   When the mass swept up by the remnant becomes comparable to the mass of the
   ejecta, the evolution switches to an adiabatic phase as described by the
   Sedov-Taylor model \citep{Sedov59} where the shell radius evolves with time
   as $R_s \propto t^{2/5}$. This phase lasts until the gas behind the shock
   cools down efficiently (which typically happens around $10^6\ \mathrm{K}$),
   at which point a dense shell forms and snowplows through the ISM thanks to
   the pressure of the hot gas inside. The snowplow phase has two stages: at the
   beginning, the evolution is driven by the internal pressure, following the
   pressure-driven snowplow model \citep{Cioffi88}. If the ambient medium is
   dense, this phase can be very short and have a timescale similar to the
   transition timescale. When the internal pressure support vanishes, the shell
   expands freely through the medium, following the momentum-conserving snowplow
   model \citep{Oort51} where the shell radius follows $R_s \propto t^{1/4}$. To
   address the question of the efficiency with which supernovae can drive the
   turbulence into the galaxies, an important question is how much momentum is
   delivered by supernova remnants in the ISM? In this respect a remarkable
   result has been inferred \citep[e.g.,][]{blondin+1998}: the total momentum
   injected by the supernova has a weak dependence on the gas density (see below
   for more details) of the surrounding gas suggesting that a constant value can
   be assumed in simulations.

   Generally speaking, because the ISM is very inhomogeneous, the medium in
   which supernova remnants propagate is expected to present density contrasts
   of several orders of magnitude. How does it modify the supernova remnant
   evolution? Does it affect the total amount of momentum that is eventually
   delivered in galaxies? Another related aspect that has received little
   attention is the influence that magnetic fields can have on the remnant, in
   particular whether the magnetic field can alter the amount of momentum
   injected into the ISM.

   A complementary question is the impact that supernovae can have on molecular
   clouds as they explode. While the impact of HII regions \citep{matzner2002,
   dale+2012, dale+2013}, protostellar outflows \citep{li-nakamura2006}, and the
   combined effect of protostellar outflows and stellar winds \citep{dale+2014}
   has been investigated, the impact supernovae might have on the molecular
   clouds has received less attention.  However, it is currently estimated that
   about $10$ -- $20\,\%$ of the observed supernova remnants interact with a
   molecular cloud \citep{fukui2003,hewitt2009,chen2013a}. The significance of
   this number is, however, not straightforward. While one would expect about
   $1000$ supernova remnants in our Galaxy, only $\simeq 300$ have been observed
   so far \citep[e.g.,][]{brogan2006}. The reason for this difference has not
   been clearly established. On the one hand supernovae that explode in a
   rarefied medium ($n \ll 1$ cm$^{-3}$) may explode without leaving an
   identifiable remnant \citep{chu1990}, but on the other hand, supernovae that
   explode in a dense medium will have a small remnant radius and will quickly
   dissipate (see Fig.~\ref{graph-momentum-uniform}), and thus may also be
   difficult to see. Therefore the question of where most of the remnants are
   located remains to be clarified.

   Motivated by evidence for interaction between supernovae and molecular clouds
   such as N49 \citep{shull1983}, a few analytical studies have been performed
   \citep{shull1985,chevalier1999}. More recently, hydrodynamical numerical
   simulations of a supernova explosion taking place in a dense clump have been
   performed by \citet{rogers2013}. These simulations include stellar winds,
   which operate before supernovae explode. They do not however include gravity
   and magnetic field. The authors conclude that the supernova remnant tends to
   escape the cloud quickly without depositing a substantial fraction of its
   energy. They nevertheless show that the supernovae trigger a mass flux that
   is about 10 times the loss induced by the stellar winds.

   In this paper, we investigate the evolution of supernova remnants in
   molecular clouds. To quantify accurately the amount of momentum that is
   delivered in the surrounding medium, we focus on a single event. We perform
   numerical simulations of supernovae in uniform and turbulent, cloud-like
   medium, taking into account cooling, self-gravity, and magnetic field. We
   specifically investigate the role of the last by comparing hydrodynamical and
   magnetized simulations. To assess our calculations, we present a simple but
   robust model for momentum and kinetic energy feedback, based on the
   well-known remnant evolution models \citep{Oort51, Sedov59} and compare it to
   our simulations.

   Section \ref{sect-num} describes the numerical setup: the physics taken into
   account, the initial conditions, and the resolution. The results of the
   uniform medium simulations are detailed in Sect.~\ref{sect-unif}, and the
   results of the turbulent simulations are described in Sect.~\ref{sect-turb}.
   Section \ref{sect-concl} concludes the paper.

%__________________________________________________________________

\section{Numerical setup}\label{sect-num}

   \subsection{Physical processes}

      Our simulations include various physical processes known to be important
      in molecular clouds. We solve the ideal magnetohydrodynamics (MHD)
      equations with self-gravity and take into account the cooling and heating
      processes relevant to the ISM.

      The equations we solve are
      \begin{align}
         \partial_t \rho + \vec{\nabla} \cdot \left( \rho \vec{v} \right) &= 0, \\
         \partial_t \left( \rho \vec{v} \right)
            + \vec{\nabla} \cdot \left( \rho \vec{v} \otimes \vec{v}
               + \left( P + \frac{B^2}{8\pi} \right) \tens{I}
               - \frac{\vec{B} \otimes \vec{B}}{4\pi} \right)
            &= -\rho\vec{\nabla} \phi, \\
         \partial_t E
            + \vec{\nabla} \cdot \left(\left( E + P - \frac{B^2}{8\pi} \right) \vec{v}
               + \frac{1}{4\pi} \vec{B} \times \left( \vec{v} \times \vec{B} \right) \right)
            &= -\rho \vec{v} \cdot \vec{\nabla} \phi - \rho\mathcal{L}, \\
         \partial_t \vec{B} - \vec{\nabla} \times \left( \vec{v} \times \vec{B} \right) &= 0, \\
         \Delta \phi - 4\pi G \rho &= 0,
      \end{align}
      with $\rho$, $\vec{v}$, $P$, $\vec{B}$, $\phi$, and $E$ respectively being
      the density, velocity, pressure, magnetic field, gravitational potential,
      and total (kinetic plus thermal plus magnetic) energy. The loss function
      $\mathcal{L}$, includes UV heating and a cooling function with the same
      low-temperature part as in \citet{ah05} and the high-temperature part
      based on \citet{sutherland+1993}, resulting in a function similar to the
      one used in \citet{joung+2006}.

      We trigger supernovae at given positions in space and time by injecting
      $10^{51}\ \mathrm{erg}$ of thermal energy within a sphere of radius equal
      to 2 computing cells.

%__________________________________________________________________

   \subsection{Uniform density simulations}

      As a preliminary study, we run simulations of a supernova remnant going
      through a uniform medium, in order to compare our results with previous
      work \citep{Oort51, Sedov59, Chevalier74, Cioffi88, Ostriker88}. We
      consider several densities ranging from $1$ to $1000$ particles per cubic
      centimeter, with an appropriate box size between $160\ \mathrm{pc}$ and
      $40\ \mathrm{pc}$. The gas is initially at thermal equilibrium. Table
      \ref{table-uniform-ic} gives the four initial conditions.

      \begin{table}
         \begin{center}
            \begin{tabular}{ccc}
               \hline\hline
               Box size ($\mathrm{pc}$) & Density ($\mathrm{cm}^{-3}$)
                  \rule{0pt}{2ex} & Temperature ($\mathrm{K}$) \\
               \hline
               160 & 1    & 4907.8 \\
               80  & 10   & 118.16 \\
               80  & 100  & 36.821 \\
               40  & 1000 & 19.911 \\
               \hline
            \end{tabular}
         \end{center}
         \caption{Initial conditions for the uniform cases. The gas is
            initially at rest.}
         \label{table-uniform-ic}
      \end{table}

      To study the influence of the magnetic field, we run a set of
      hydrodynamical simulations and a set of MHD simulations for which the
      magnetic field is initially uniform and has an intensity of $5\
      \mathrm{\mu G}$. The results of these simulations are similar to the
      hydrodynamical case and are described in Appendix~\ref{sect-mag}.

%__________________________________________________________________

   \subsection{Turbulent simulations}
      Clouds are turbulent and present large density contrasts and a complex
      structure induced by supersonic turbulence. In order to reproduce this in
      our simulations, we set up a spherical density profile and we add a
      turbulent velocity field; we then let the cloud evolve for about one
      crossing time.

      More precisely, we consider a spherical cloud with a profile given by
      $\rho = \rho_0 / (1+(r/r_0)^2)$ where $\rho_0=9370\ \mathrm{cm}^{-3}$ and
      $r_0 = 1.12\ \mathrm{pc}$.  The edge density is initially equal to
      $\rho_0/10$.  The total mass within this inner region is $10^4\
      \mathrm{M_\odot}$. In order to avoid a sharp transition with the diffuse
      ISM and to mimic the HI haloes observed around molecular clouds
      \citep[e.g.,][]{elmegreen+1987, hf2012}, we add around it a uniform
      density cloud of density $\rho_0/100 \simeq 93\ \mathrm{cm^{-3}}$. The
      radius of this HI haloes is about $6.8\ \mathrm{pc}$ and it contains a
      mass equal to about $5 \times 10^3 \ \mathrm{M_\odot}$. In the rest of the
      computational box the initial density is equal to $1\ \mathrm{cm^{-3}}$.
      The cloud is initially at thermal equilibrium and is surrounded by warm
      neutral medium at a temperature of $8000\ \mathrm{K}$.

      The ratio of the thermal over gravitational energy is initially equal to
      about $1\,\%$. We inject a turbulent velocity field into the cloud. The
      velocity field presents a Kolmogorov power spectrum and has random phase.
      The kinetic energy is initially about $100\,\%$ of the gravitational
      energy meaning that the cloud is globally supported by turbulence. We let
      the cloud evolve for $1.25\ \mathrm{Myr}$ in order for the turbulence to
      develop and trigger density (and magnetic field) fluctuations
      self-consistently, before adding a supernova.

\begin{figure}
         \begin{center}
            \includegraphics[width=80mm]{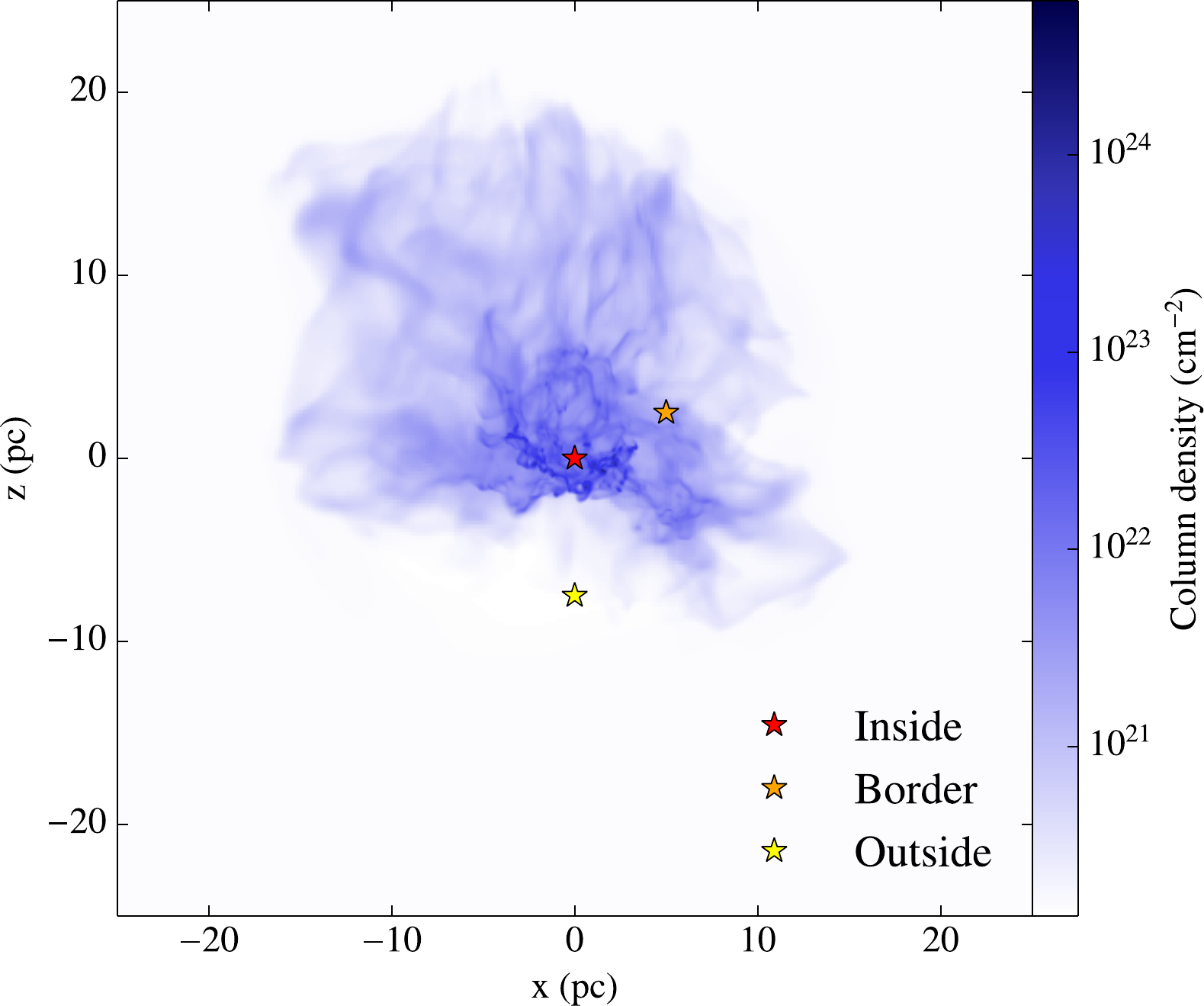}
         \end{center}
         \caption{Positions of the supernovae injected in the cloud simulations.}
         \label{graph-sn-pos}
      \end{figure}

      To explore the influence that the supernova position may have on the
      result, we run three simulations with three different supernova positions
      (inside, at the border and outside of the cloud), and one without
      supernova. Figure~\ref{graph-sn-pos} shows the positions of the supernova
      explosions with respect to the cloud. In the inside run, the supernova
      explodes in a density of about $700\ \mathrm{cm^{-3}}$, in the border run
      the density is closer to $20\ \mathrm{cm^{-3}}$, and in the outside run it
      is about $1.2\ \mathrm{cm^{-3}}$.

%__________________________________________________________________

   \subsection{Numerical code and resolution}

      We run our simulations with the RAMSES code \citep{teyssier2002,
      fromang+2006}, an adaptive mesh code using a Godunov scheme and a
      constrained transport method to solve the MHD equations, therefore
      ensuring the nullity of the magnetic field divergence. We use two levels
      of adaptive mesh refinement on a $256^3$ base grid, leading to a maximum
      resolution around $0.05\ \mathrm{pc}$ for the turbulent case for which the
      box size is equal to $50\ \mathrm{pc}$, and between $0.04$ and $0.16\
      \mathrm{pc}$ for the uniform case. The refinement criterion is based on
      the Jeans length, which must be described by at least 10 computational
      cells in the turbulent cloud runs and on the pressure gradient in the
      uniform density runs. We limited the resolution because of the presence of
      very hot (temperatures over $10^7\ \mathrm{K}$), high-velocity (over $100\
      \mathrm{km}\ \mathrm{s}^{-1}$) gas due to the supernova, which enforces a
      small time step. We integrate until we reach stationarity in the uniform
      case. In the turbulent cloud runs, since gravity is treated, the cloud
      keeps evolving and we have integrated up to about $5\ \mathrm{Myr}$ which
      is sufficient to quantify the impact of the supernova on the cloud.

      For the turbulent simulations, the blast waves quickly reach the edge of
      the computational domain and in order to let the energy flow out of the
      box, we use vanishing gradient boundary conditions, except for the normal
      component of the magnetic field for which a vanishing divergence condition
      is enforced.

      To assess our results, we have also performed two runs with a base grid of
      $512^3$ and two more adaptive mesh refinement levels. One of these runs
      has the same computational box size of $50\ \mathrm{pc}$, therefore
      implying a spatial resolution 2 times higher while the other one has a box
      size equal to $100\ \mathrm{pc}$, which allows us to follow the supernova
      remnant for a longer time and to verify that the total amount of momentum
      injected in the surrounding ISM has indeed reached stationarity.

%__________________________________________________________________

\section{Evolution of a supernova remnant in a uniform medium}\label{sect-unif}

   As recalled previously, three phases are typically expected during supernova
   explosions, the free expansion phase for which the mass of the remnant
   dominates the swept-up mass; the Sedov-Taylor phase, during which the
   expansion is adiabatic; and finally the snowplow phase, during which the
   dense shell can radiate energy efficiently. In our simulations, given the
   temporal and spatial resolution, we cannot observe the free expansion phase.
   We observe the adiabatic phase, the pressure-driven snowplow phase, and the
   momentum-conserving snowplow phase (for the runs with the highest ambient
   densities).

   \subsection{Simple analytical trends}
      \label{analyt}
      For the adiabatic phase, Sedov's analytical model \citep{Sedov59} predicts
      a total momentum of
      \begin{equation}
         p_{43} = 1.77 n_0^{1/5} E_{51}^{4/5} t_4^{3/5},
         \label{eq-momentum}
      \end{equation}
      where $p_{43}$ is the total momentum in units of
      $10^{43}~\mathrm{g~cm~s^{-1}}$, $n_0$ is the particle density in
      $\mathrm{cm}^{-3}$, $E_{51}$ is the supernova energy in units of
      $10^{51}~\mathrm{erg}$, and $t_4$ is the age of the remnant in units of
      $10^4~\mathrm{yr}$.

      We define the transition time $t_{tr}$ as the moment when the age of the
      remnant becomes equal to the cooling time $\tau_{cool}$ of the shell which
      is given by
      \begin{equation}
         \tau_{cool} = \frac{3}{2} k_B \frac{n_s T_s}{n_s^2 \Lambda_s},
         \label{cool_time}
      \end{equation}
      where $n_s$ and $T_s$ are the gas density and temperature of the shell,
      and $\Lambda_s$ the net cooling (in $\mathrm{erg}\ \mathrm{cm}^{3}\
      \mathrm{s}^{-1}$).

      Below we estimate the transition time $t_{tr}$ numerically, but it is
      worth inferring explicitly the relevant dependence. The net cooling
      $\Lambda_s$ can be approximated as \citep[e.g.,][]{blondin+1998}
      \begin{equation}
         \Lambda (T) \propto {1 \over T}.
         \label{cool_simp}
      \end{equation}
      The temperature and the density in the shell are given by the
      Rankine-Hugoniot conditions. For a monoatomic gas we thus have
      \begin{equation}
         n_s = 4 n_0, \\
         T_s \propto v_s^2.
         \label{dens_temp}
      \end{equation}
      The evolution of the shell radius is given by $R_s \propto (E
      t^2 / n_0)^{1/5}$. Thus we get
      \begin{equation}
         v_s^4 \propto n_0^{-4/5} E^{4/5} t^{-12/5}.
         \label{velo_shell}
      \end{equation}
      Combining Eqs.~(\ref{cool_time}),~(\ref{cool_simp}),~(\ref{dens_temp}),
      and~(\ref{velo_shell}), we get
      \begin{equation}
         t_{tr} \simeq \tau_{cool} \propto \frac{T_s^2}{n_s} \propto n_0^{-9/5} E^{4/5} t_{tr}^{-12/5},
         \label{time_eq1}
      \end{equation}
      therefore
      \begin{equation}
         t_{tr} \propto n_0^{-9/17} E^{4/17}.
         \label{time_eq2}
      \end{equation}
      Using Eq.~(\ref{eq-momentum}) and Eq.~(\ref{time_eq2}), we can estimate
      the dependence of the shell momentum at the transition time, $t_{tr}$, and
      we get
      \begin{equation}
         p_{43} \propto n_0^{-2/17} E^{16/17}.
         \label{time_eq3}
      \end{equation}

      Thus, we see that the total momentum delivered in the ISM has a weak
      dependence on the medium density. For example, changing the density by a
      factor of $10^3$, leads to a momentum variation of about $2$ -- $2.5$.

%__________________________________________________________________

   \subsection{Momentum injection: result}
      \label{uni_explos}
\begin{figure}
         \begin{center}
            \includegraphics[width=80mm]{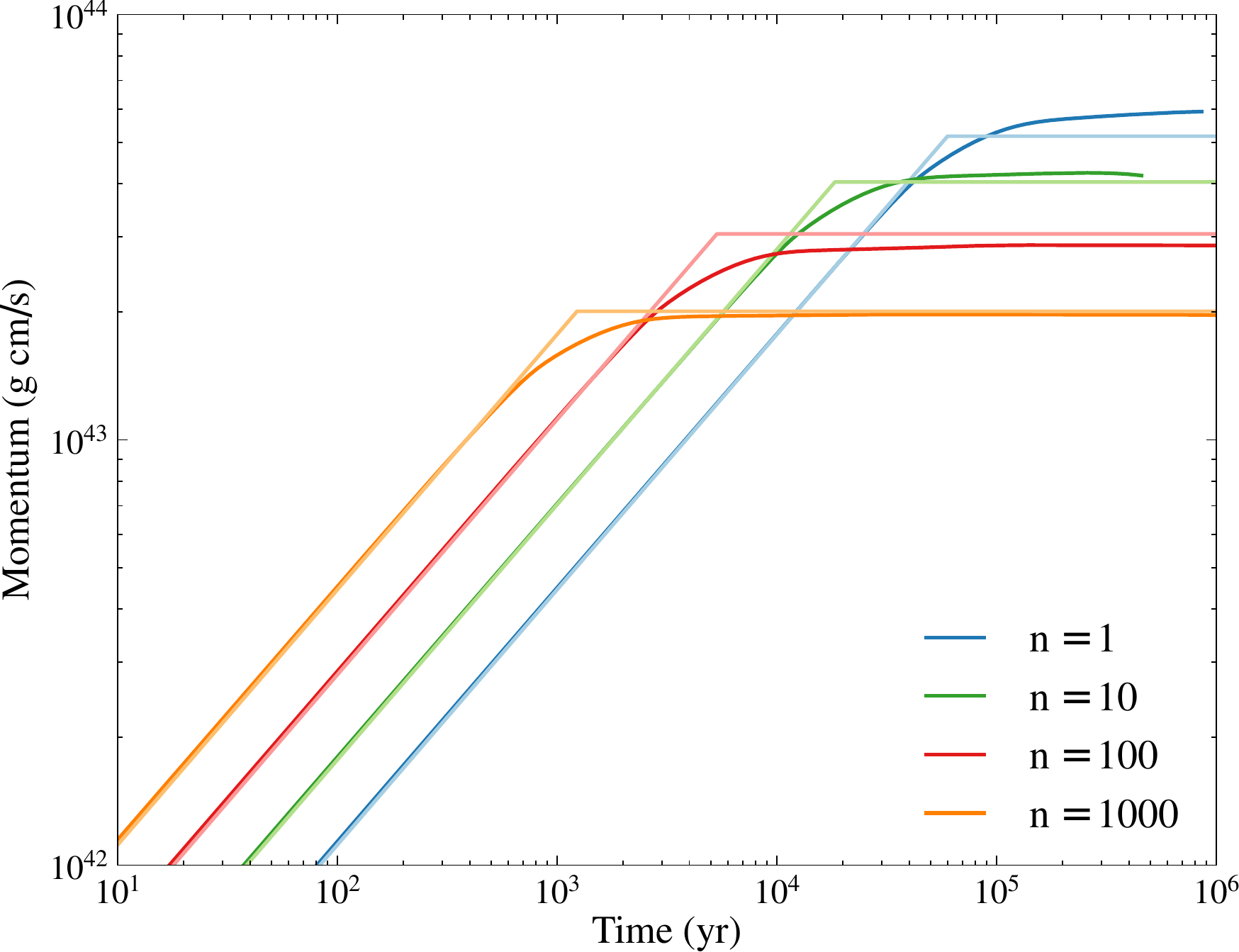}
         \end{center}
         \caption{Integrated radial momentum for the uniform simulations.  The
            thin straight lines correspond to the analytical trends described in
            Sect.~\ref{analyt}. Two main phases can be distinguished. First,
            during the Sedov-Taylor phase the momentum increases with time.
            Then, during the radiative snowplow phase the momentum is nearly
            constant. The dependence of the momentum on the ISM density is
            rather shallow.
         }
         \label{graph-momentum-uniform}
      \end{figure}

      Figure~\ref{graph-momentum-uniform} shows the integrated radial momentum
      as a function of time for the four uniform density simulations. We clearly
      observe two phases of the remnant's evolution: the adiabatic
      (Sedov-Taylor) phase, conserving energy for which the momentum follows $p
      \propto n_0^{1/5} t^{3/5}$, and the momentum-conserving (for the highest
      ISM densities only), and pressure-driven snowplow \citep{Oort51,
      Cioffi88}.  Figure~\ref{graph-momentum-uniform} also shows the analytical
      prediction stated by Eq.~\eqref{eq-momentum} and valid before the
      transition time.  As can be seen the agreement is excellent.

      To model analytically the second phase, we solve numerically the equation
      $t_{tr} = \tau_{cool}$; that is to say, we use the complete cooling
      function instead of using Eq.~(\ref{cool_simp}). For the highest ambient
      densities ($n_0 \gtrsim 10$), the momentum injection is reasonably well
      fitted by the momentum-conserving snowplow model with the final momentum
      $p_f$ taken to be the momentum of a Sedov-Taylor blast wave at $2t_{tr}$
      (the numerical values are given in Table \ref{table-cooling-time}). Some
      small deviations are found with the lowest density case because the
      pressure within the shell is still higher than the surrounding pressure
      and the shell keeps accelerating. When the surrounding gas density varies
      by 3 orders of magnitude, the total momentum varies by a factor of about
      $2.5$.

      \begin{table}
         \begin{center}
            \begin{tabular}{ccc}
               \hline\hline
               $n_0$ ($\mathrm{cm}^{-3}$) & $t_{tr}$
                  ($10^4\ \mathrm{yr}$) & $p_f$
                  ($10^{43}\ \mathrm{g}\ \mathrm{cm}\ \mathrm{s}^{-1}$)
                  \rule{0pt}{2ex} \\
               \hline
               1    & 2.99   & 5.18 \\
               10   & 0.919  & 4.04 \\
               100  & 0.267  & 3.04 \\
               1000 & 0.0616 & 2.01 \\
               \hline
            \end{tabular}
         \end{center}
         \caption{Transition time $t_{tr}$ and final momentum $p_f$ as a
            function of the ambient density $n_0$.}
         \label{table-cooling-time}
      \end{table}

      We also compared the injected kinetic energy with analytical trends
      derived from the same models, again with good agreement. The details are
      given in Appendix~\ref{sect-kin}.

      Altogether, the numerical and analytical results agree well with previous
      works. Importantly, they show that the total amount of momentum delivered
      in the surrounding ISM is expected to have a weak dependence on the
      surrounding density.

%__________________________________________________________________

   \begin{figure*}
      \begin{center}
         \includegraphics[width=160mm]{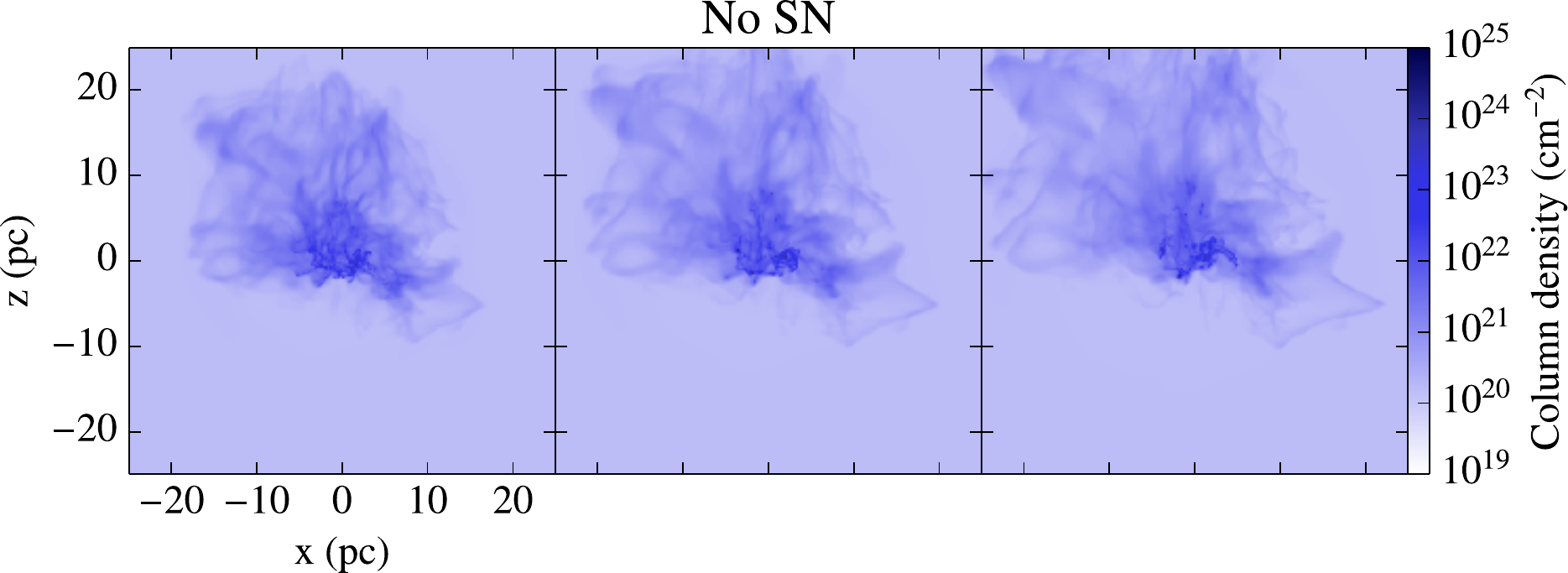}\\
         \includegraphics[width=160mm]{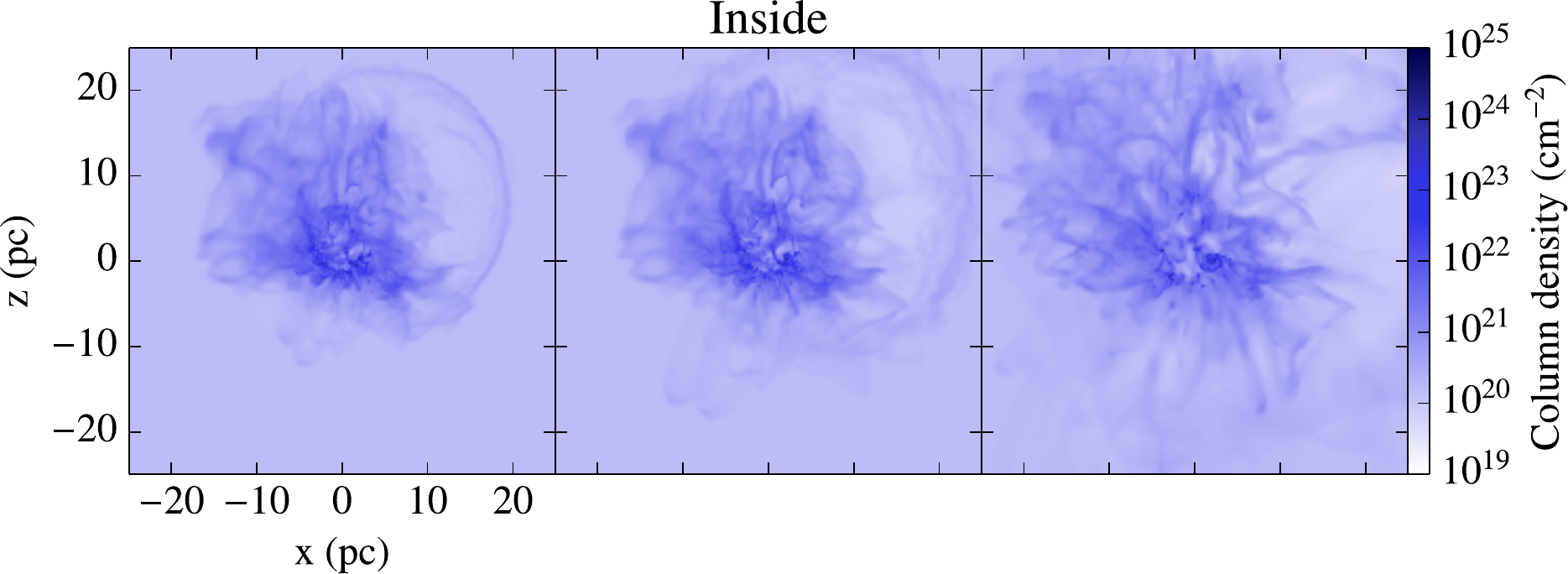}\\
         \includegraphics[width=160mm]{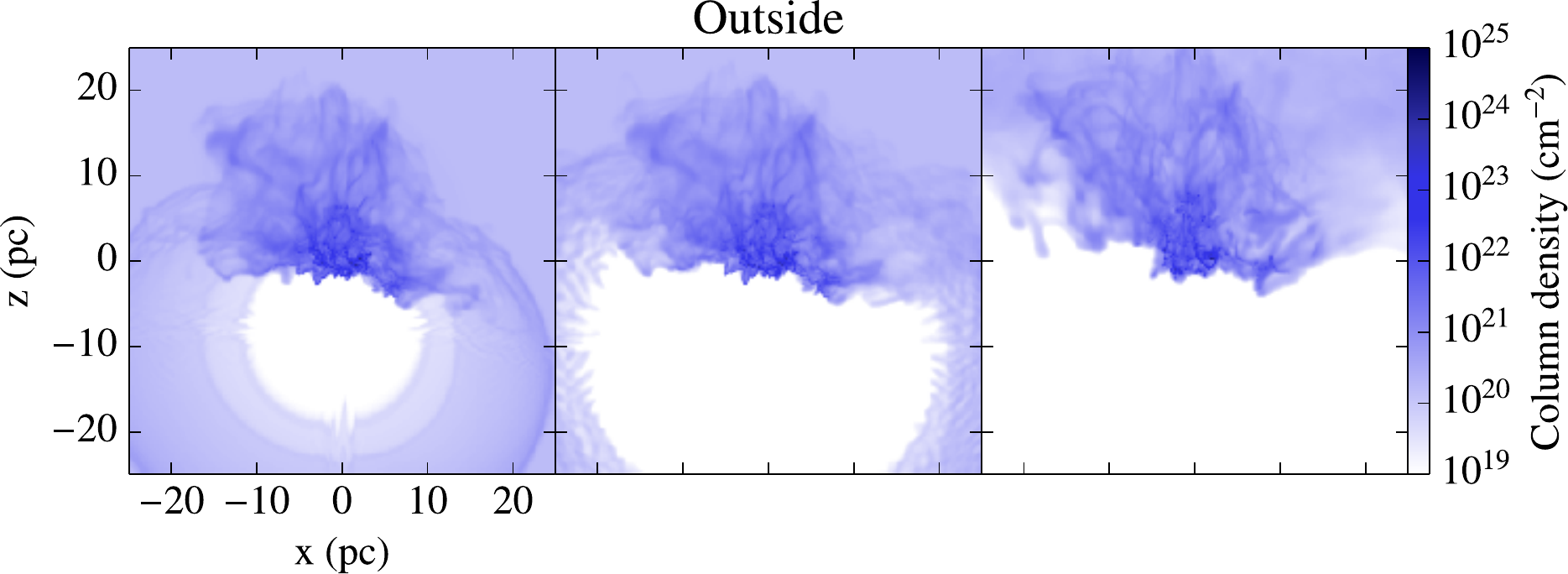}
      \end{center}
      \caption{Column density maps for the turbulent hydrodynamical simulations.
         Top panel: no supernova, $0.27, 0.85, 1.4\ \mathrm{Myr}$ after injection
         time ; middle panel: supernova inside ; bottom panel: supernova outside,
         both after $100$, $200$ and $750\ \mathrm{kyr}$.}
      \label{rt-turb}
   \end{figure*}

   \begin{figure*}
      \begin{center}
         \includegraphics[width=160mm]{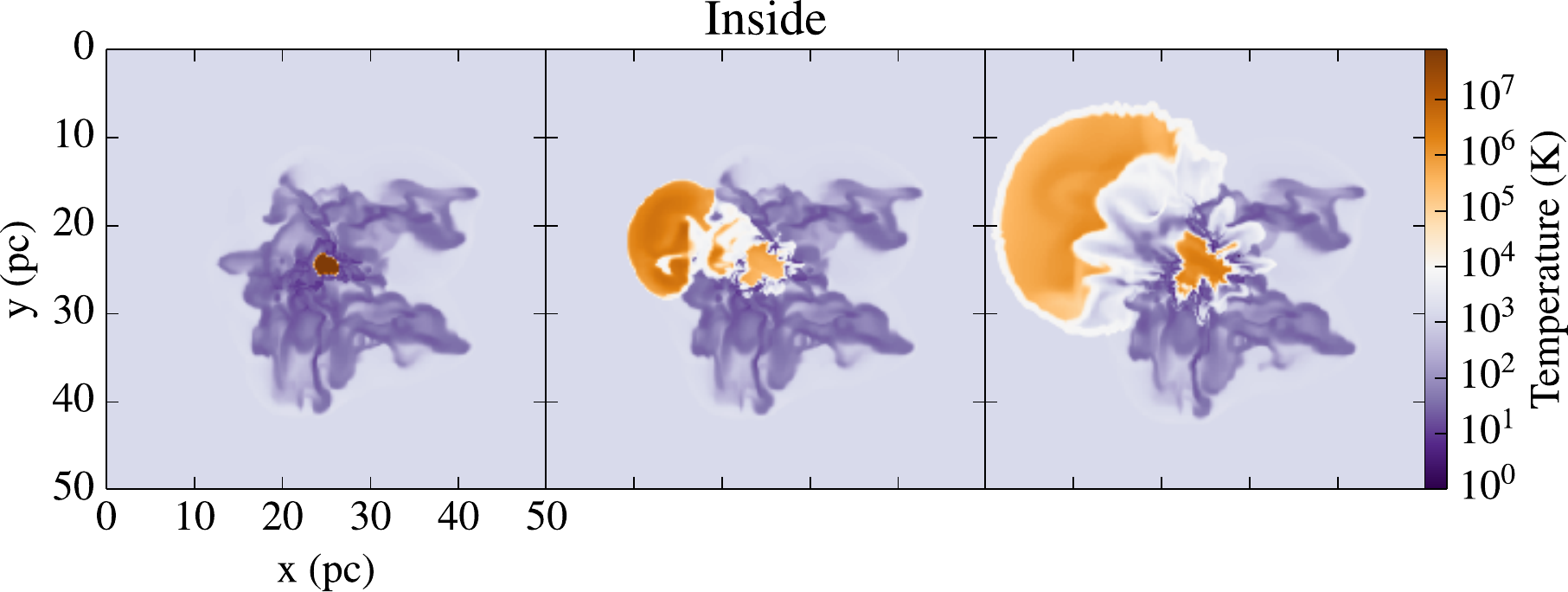}\\
         \includegraphics[width=160mm]{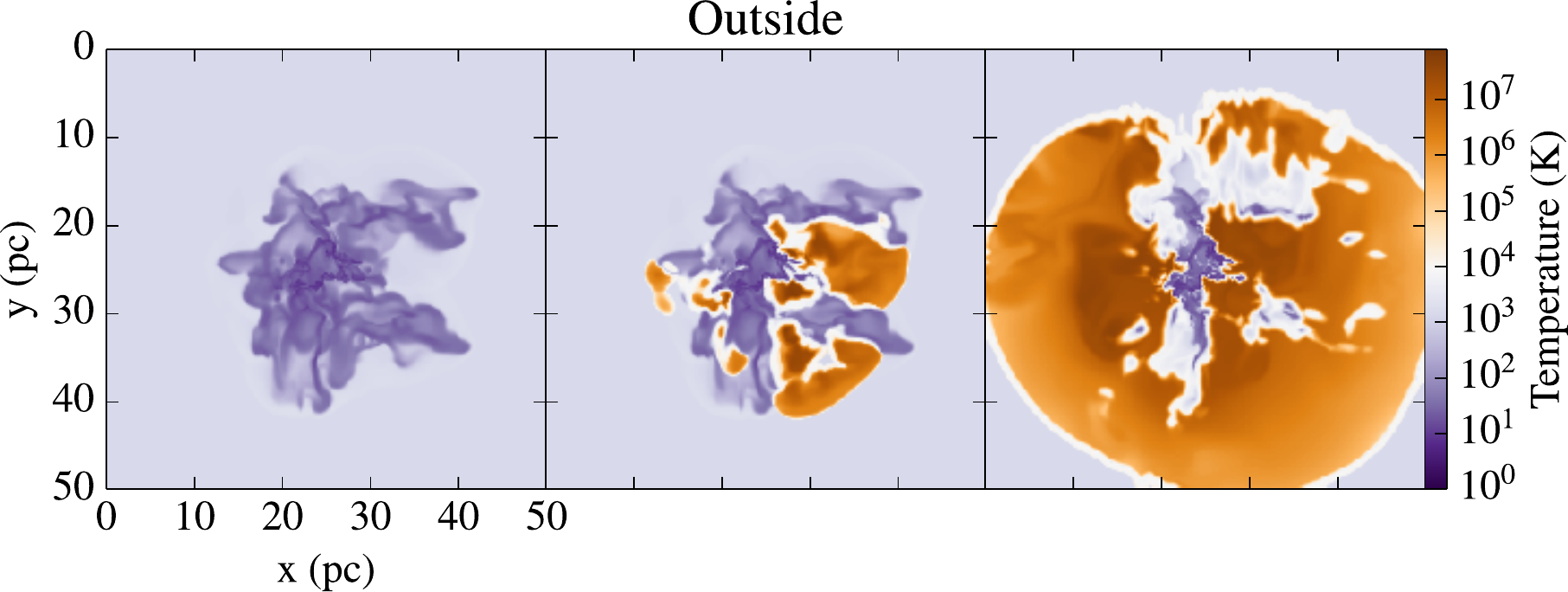}
      \end{center}
      \caption{Temperature maps for the turbulent hydrodynamical simulations
         $1$, $20$ and $100\ \mathrm{kyr}$ after supernova injection time. Top
         panel: supernova inside ; bottom panel: supernova outside (in this
         view the explosion takes place behind the cloud).}
      \label{tmap-turb}
   \end{figure*}

\section{Supernova explosions in turbulent molecular clouds}\label{sect-turb}
   We now present the simulation results for supernova explosions within
   molecular clouds. We performed both hydrodynamical and MHD runs. The results
   of the MHD runs are similar to the hydrodynamical runs and are described in
   Appendix~\ref{sect-mag}.

   \subsection{Qualitative description}

      Figures~\ref{rt-turb} and \ref{tmap-turb} show the column density and a
      temperature slice ($z = 0$) of the cloud without supernova (top panels),
      with the supernova inside (middle panels) and outside (bottom panels) (see
      Fig.~\ref{graph-sn-pos}) at three different time steps. The top panels
      show a complex multi-scale column density similar to what has been found
      in many simulations \citep[e.g., ][]{ballesteros+2007, hf2012}. The dense
      gas rapidly collapses into several objects under the influence of gravity.
      The second row shows that the hot gas that is initially located inside the
      dense cloud is rapidly able to escape through channels of more diffuse
      gas, which has been pushed by the high pressure. Once it reaches the
      outside medium, a bubble forms and propagates as found previously. The
      propagation in the rest of the cloud remains limited but happens
      nevertheless. Indeed, the high pressure enhances the contrast inside the
      cloud by creating regions of very low density and by compressing further
      the dense clumps.

      The temperature plot (Fig.~\ref{tmap-turb}) also reveals interesting
      behavior. In the inside run, the high pressure gas quickly opens up a
      chimney through the cloud by pushing the diffuse material. Once it reaches
      the diffuse ISM, the supernova remnant starts expanding and develops into
      a spherical shell. The outside run shows a different evolution. The
      explosion is broadly spherical from the very beginning. The hot gas tends
      to penetrate in between the dense regions within the cloud and quickly
      surrounds it, therefore compressing the diffuse material.

%      \begin{figure*}
%         \begin{center}
%            \includegraphics[width=160mm]{pic/turb/rt_cmp_hyd.pdf}
%         \end{center}
%         \caption{Comparison of the different supernova positions, $100\
%            \mathrm{kyr}$ after the explosion. The upper left is the case
%            without supernova, the upper right has a supernova in the center,
%            the lower left in a relatively dense zone close to the cloud
%            (border case), and the lower right in a diffuse zone close to
%         the cloud (outside case).}
%         \label{rt-comp}
%      \end{figure*}

%__________________________________________________________________

   \subsection{Total momentum injection}

      In order to quantify the overall impact a supernova embedded in a
      molecular cloud can have on the surrounding ISM, we first calculate the
      total radial momentum and kinetic energy as a function of time. Since the
      gas is initially moving, it may not be straightforward to discriminate
      between the contribution of the supernova and the initial condition.
      However, the momentum before the supernova explodes is about 10 times
      smaller than the peak value reached once the supernova has taken place.
      Because the results concerning the injection of kinetic energy are of
      lesser importance, we describe them in Appendix~\ref{sect-kin}.

      The top panel of Fig.~\ref{graph-momentum-turb} shows the evolution of
      radial momentum for the three runs, inside, border, and outside, compared
      to the models described in Sect.~\ref{uni_explos} valid for the uniform
      density runs. The vertical dashed lines show the time at which the
      supernova remnant starts leaving the box and at this point, or shortly
      after, the total momentum decreases. The bottom panel of
      Fig.~\ref{graph-momentum-turb} shows the results for the same simulation,
      but performed with a computational box 2 times larger. As can be verified
      from the total momentum values, the maximum obtained with the $50\
      \mathrm{pc}$ box is close to the maximum value obtained with the $100\
      \mathrm{pc}$ box. Numerical convergence can also been assessed from this
      same diagram as a simulation with 2 times better resolution is also
      presented.

      The injected momentum follows an evolution that is different from any of
      the uniform density models. This is relatively unsurprising given that the
      structure of the clouds is far from spherical and entails a wide
      distribution of densities which spans about 6 orders of magnitude (see
      Appendix~\ref{sect-pdf} for details). We note that the trends are
      qualitatively similar to the uniform density case; that is to say, the
      outside run tends to be closer to the lower density uniform models than
      the inside run. In spite of these significant differences with the uniform
      density runs, the total momentum injected by the supernova does not vary
      much and remains close to the values inferred in the uniform density case,
      i.e., a few $10^{43}\ \mathrm{g\ cm\ s^{-1}}$.

      This result constitutes an important generalization of the uniform models
      and suggests that the feedback from supernovae in star forming regions
      and, more generally in galaxies, can be modeled by adopting a simple
      prescription of a few $10^{43}\ \mathrm{g\ cm\ s^{-1}}$ being released in
      the ISM.

      \begin{figure}
         \begin{center}
            \includegraphics[width=8cm]{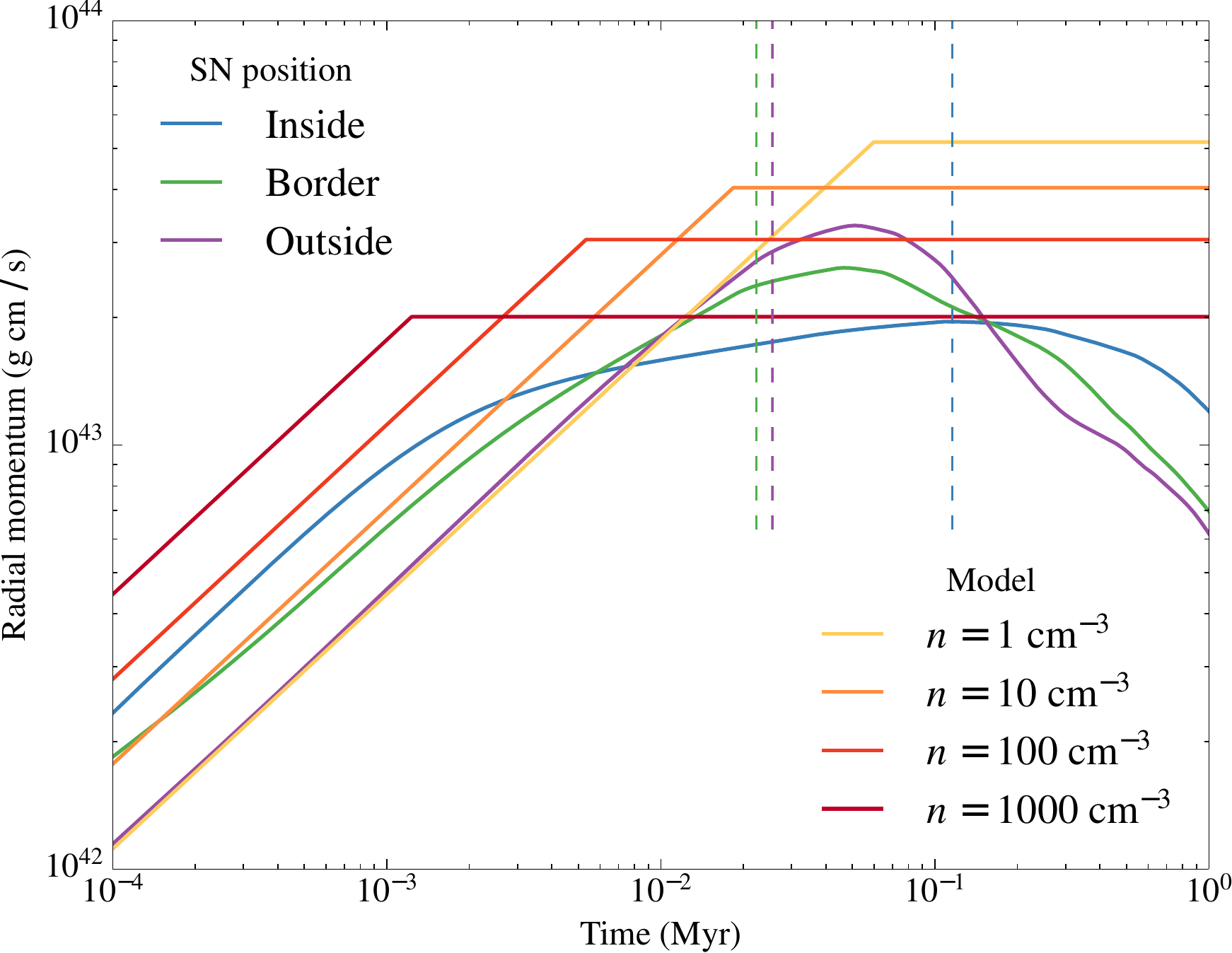}\\
            \includegraphics[width=8cm]{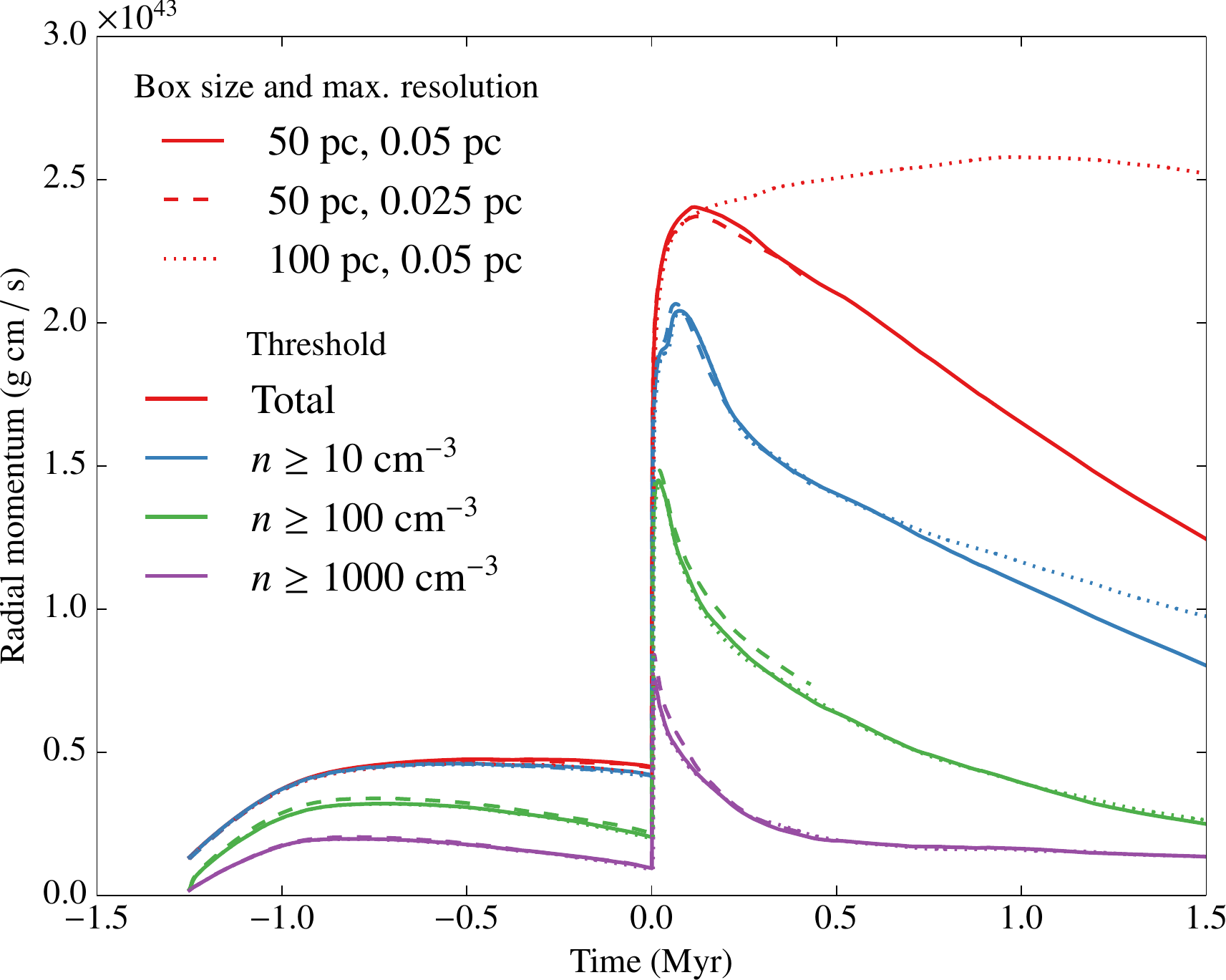}
         \end{center}
         \caption{Top panel: total radial momentum, comparison between turbulent
            simulations and the analytical model for various densities. The
            vertical lines correspond to the time at which the gas starts
            leaving the computational domain (from left to right: border,
            outside, inside).  Bottom panel: inside run for two different box
            sizes and two numerical resolutions.}
         \label{graph-momentum-turb}
      \end{figure}

%__________________________________________________________________

   \subsection{Mass distributions: impact of the supernova on the cloud}

\begin{figure}
         \begin{center}
            \includegraphics[width=75mm]{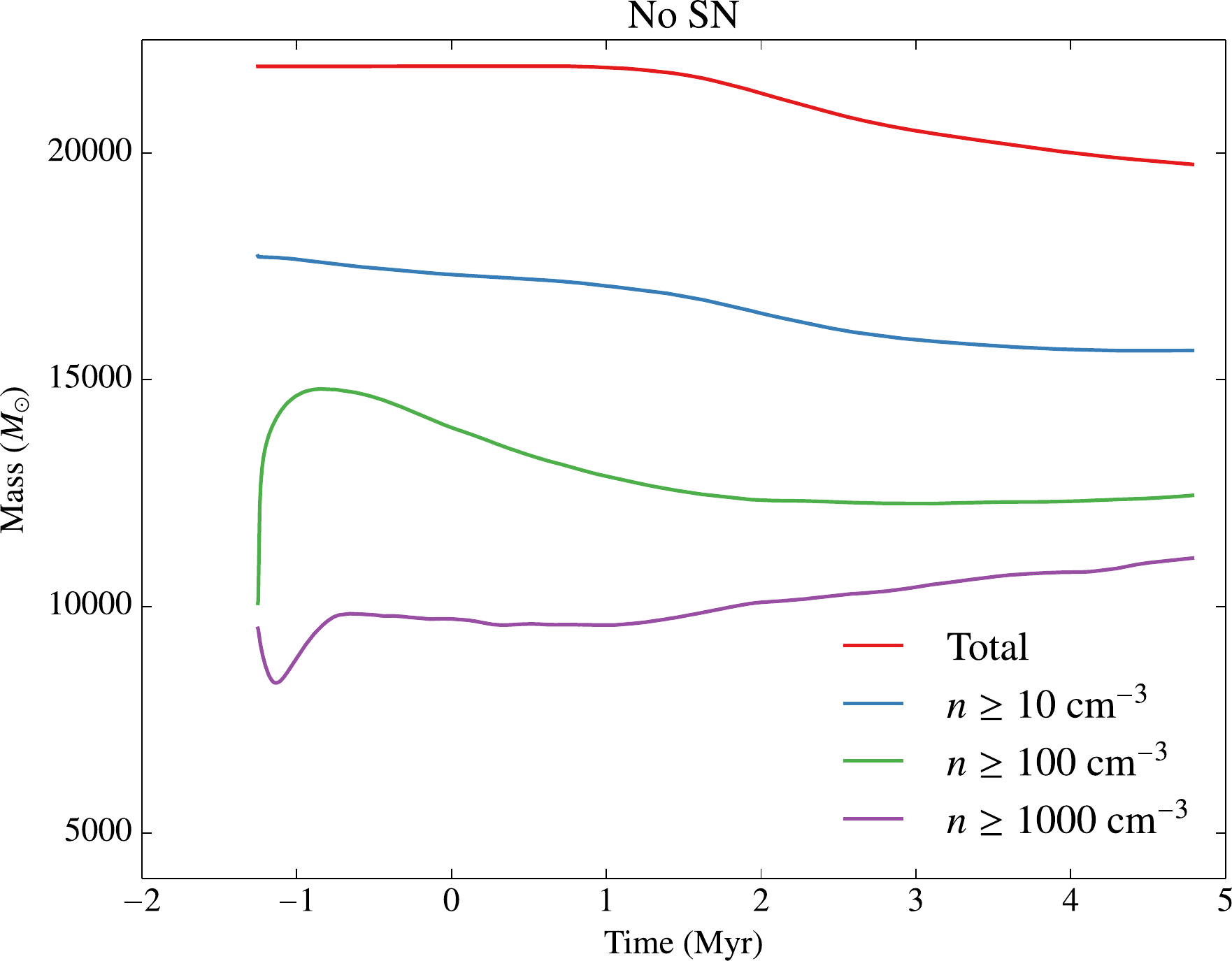}\\
            \includegraphics[width=75mm]{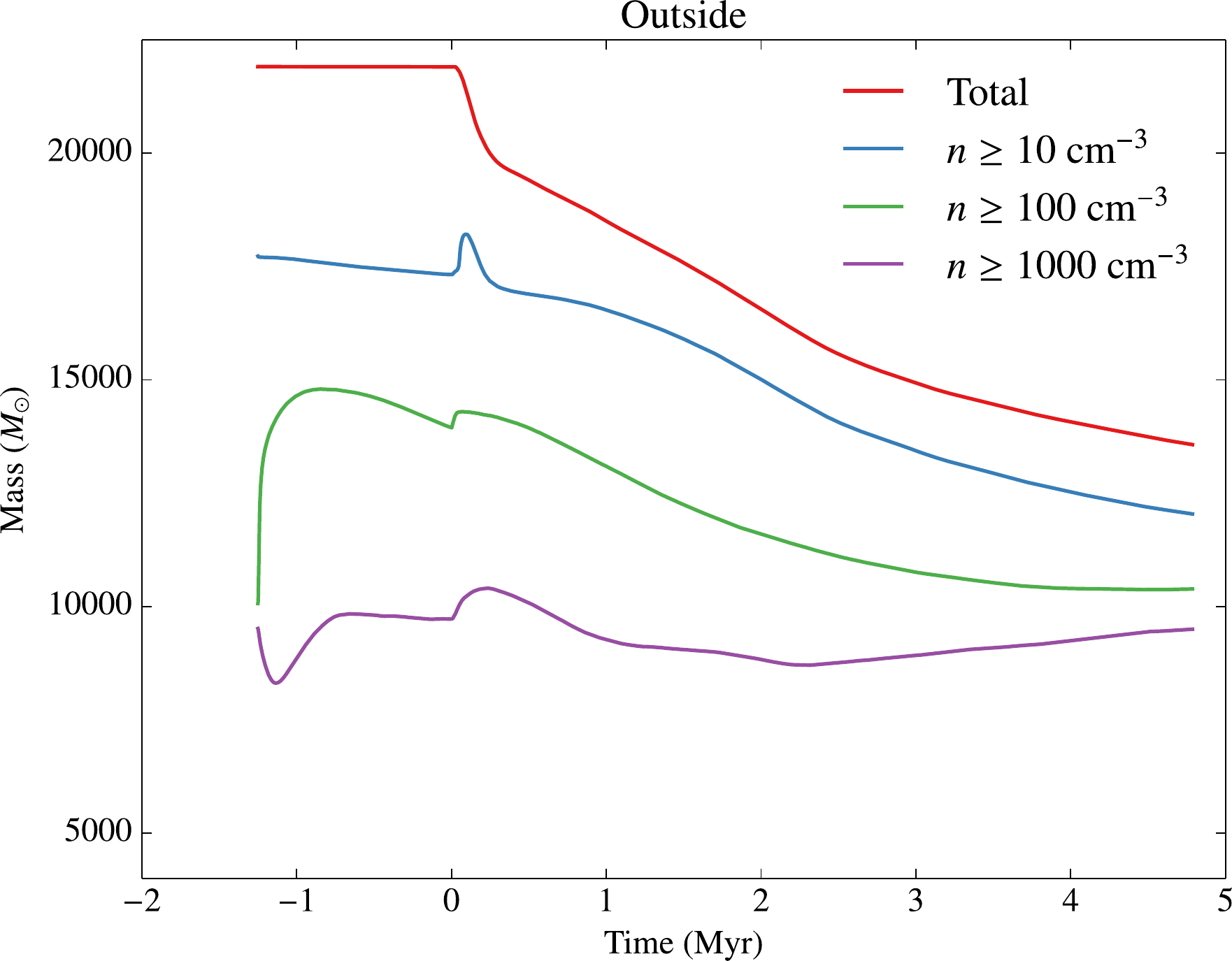}\\
            \includegraphics[width=75mm]{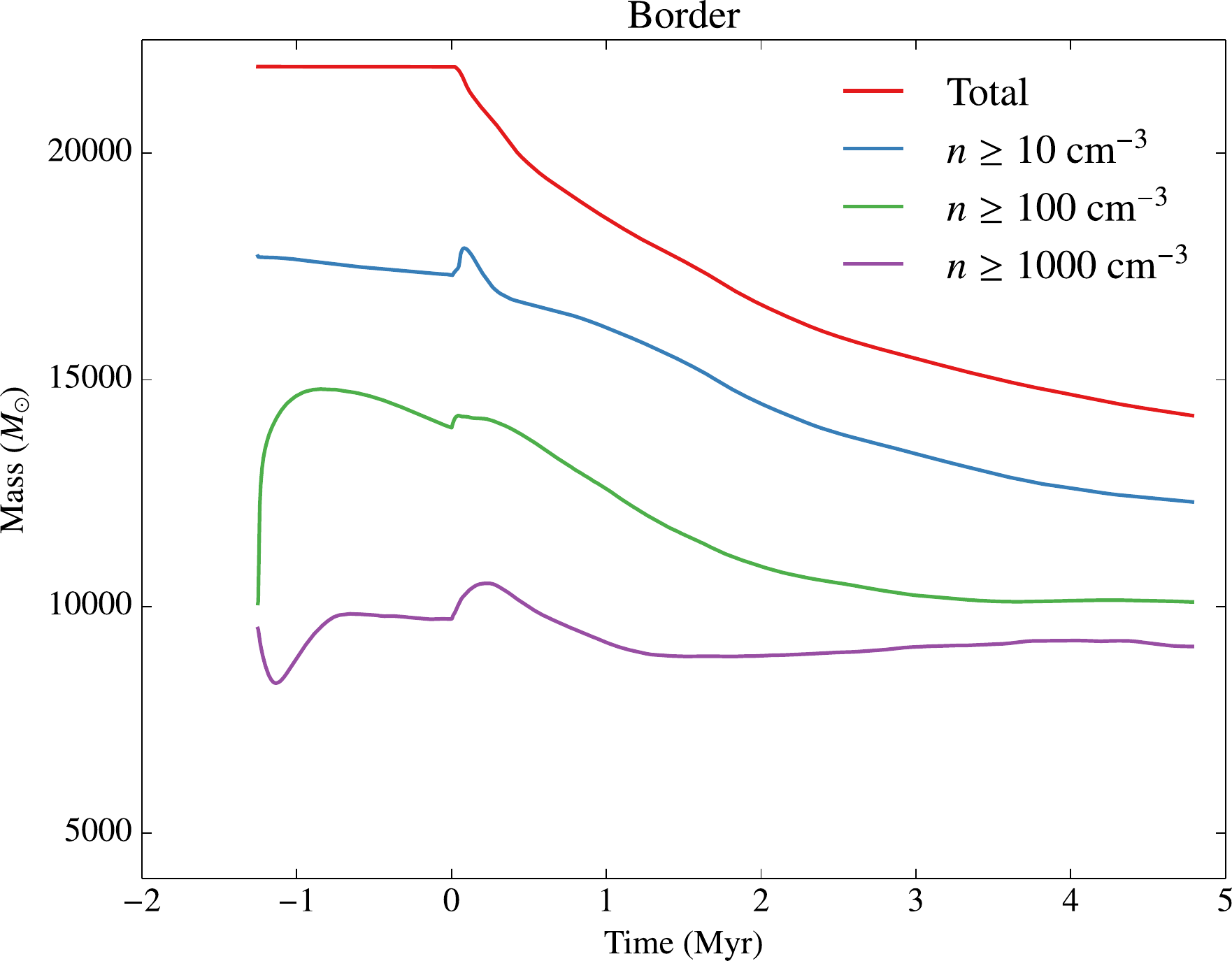}\\
            \includegraphics[width=75mm]{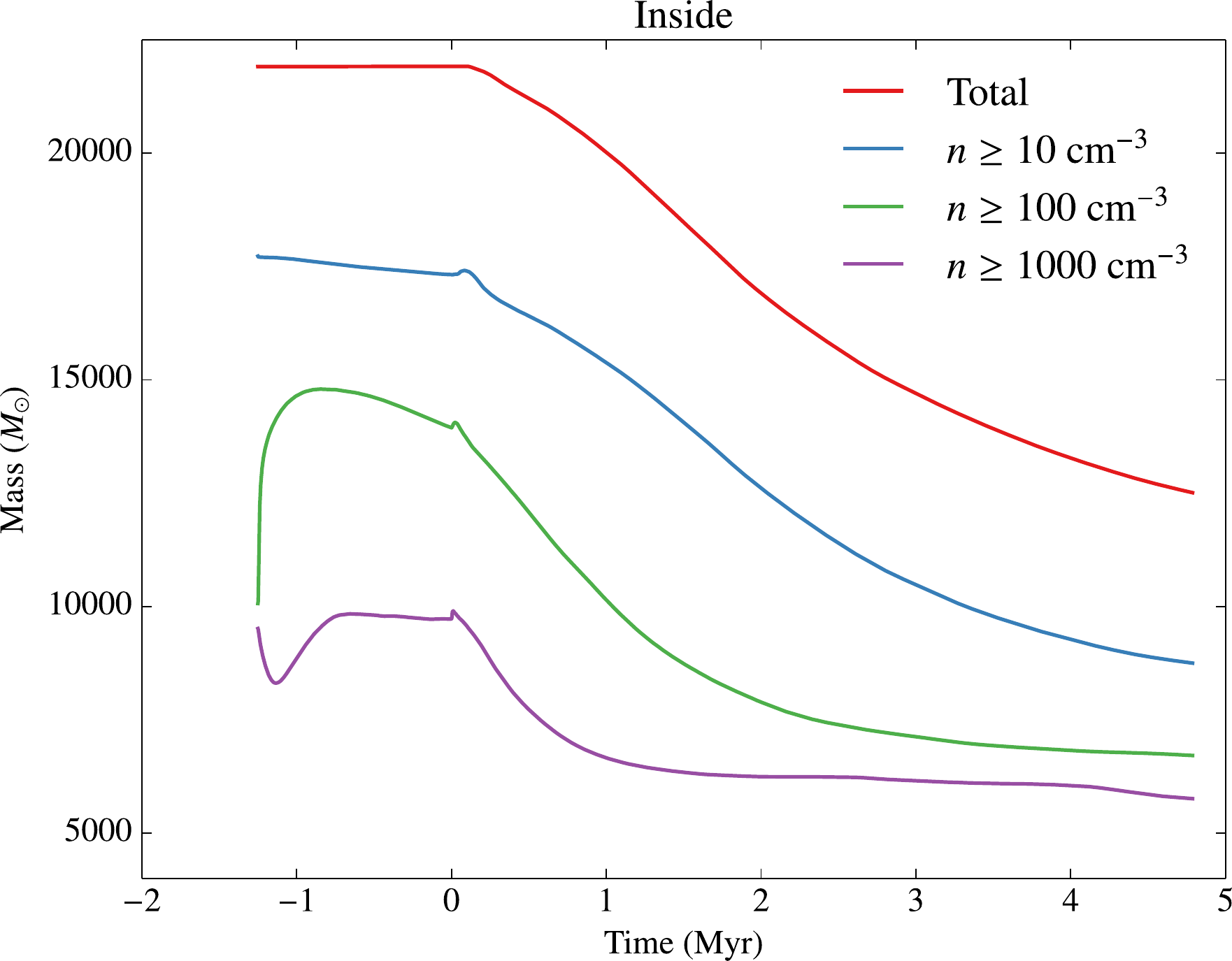}
         \end{center}
         \caption{Mass above density thresholds $10, 100, 1000\ \mathrm{cm}^{-3}$
            in the case without supernova (top panel) and outside (second
            panel), border (third panel) and inside (bottom panel)
            runs.}
         \label{graph-mass-thr}
      \end{figure}

      As can be seen in Fig.~\ref{graph-mass-thr}, which shows the mass above
      various thresholds as a function of time in the four runs (without
      supernova, outside, border, and inside), the supernova does not change the
      high-density part of the mass distribution (compared to the rest of the
      gas), but produces very diffuse gas and hot material. The details of the
      density distribution, not shown here for conciseness, are presented in
      Appendix~\ref{sect-pdf}.

      In the run without feedback (top panel), the fraction of gas of densities
      larger than 10 and $10^2\ \mathrm{cm^{-3}}$ drops during the first 2 Myr
      because the cloud is slightly supervirial and therefore expands. As
      turbulence decays, gravity takes over and the mass of gas denser than
      $10^2$ and $10^3\ \mathrm{cm^{-3}}$ increases with time as the collapse
      proceeds. We note that the decrease of the total mass is due to diffuse
      gas leaving the computational box. While this leak of material remains
      limited in the run without supernova, it is more important for the runs
      with supernova explosions. As shown in the bottom panel of
      Fig.~\ref{graph-momentum-turb}, this effect does not affect the $10^2$ and
      $10^3\ \mathrm{cm^{-3}}$ thresholds, and affects the $10\
      \mathrm{cm^{-3}}$ threshold only marginally. This can be seen by comparing
      the solid lines ($50\ \mathrm{pc}$ box size) with the dotted ones ($100\
      \mathrm{pc}$ box size).

      In the inside run (bottom panel of Fig. \ref{graph-mass-thr}), the amount
      of gas denser than $10^2$ and $10^3\ \mathrm{cm^{-3}}$ drops sharply after
      the supernova explosions and after $2\ \mathrm{Myr}$ reaches values of
      about $6$ and $7 \times 10^3\ \mathrm{M_\odot}$, respectively. Later
      evolution suggests that these numbers do not evolve significantly. At $4$
      -- $5\ \mathrm{Myr}$, these masses are almost 2 times smaller than in the
      run without supernova.  This clearly shows the impact that supernova
      feedback has on molecular clouds. For the case of the present
      configuration, that is to say a cloud of mass $\simeq 10^4\
      \mathrm{M_\odot}$ which is about 2 times supervirial, a supernova can
      reduce the mass that would eventually form stars by a factor of about 2
      (see Sect. \ref{sect-mass-ej} for a more quantitative estimate).

      As can be seen from the outside and border runs (second and third panels,
      respectively), this effect is very sensitive to the position of the
      supernova in the cloud. The amount of dense gas is only slightly reduced
      in the outside run with respect to the run without supernova. This is in
      good agreement with the results inferred by \citet{hennebelle-iffrig2014}
      where kpc simulations were performed. In particular, they found that the
      impact that supernovae have in reducing the star formation rate decreases
      as the distance between the supernova and the collapsing regions
      (represented by Lagrangian sink particles) increases.

%__________________________________________________________________

   \subsection{Momentum injection with respect to density}
      \label{sect-thr}

\begin{figure}
         \begin{center}
            \includegraphics[width=75mm]{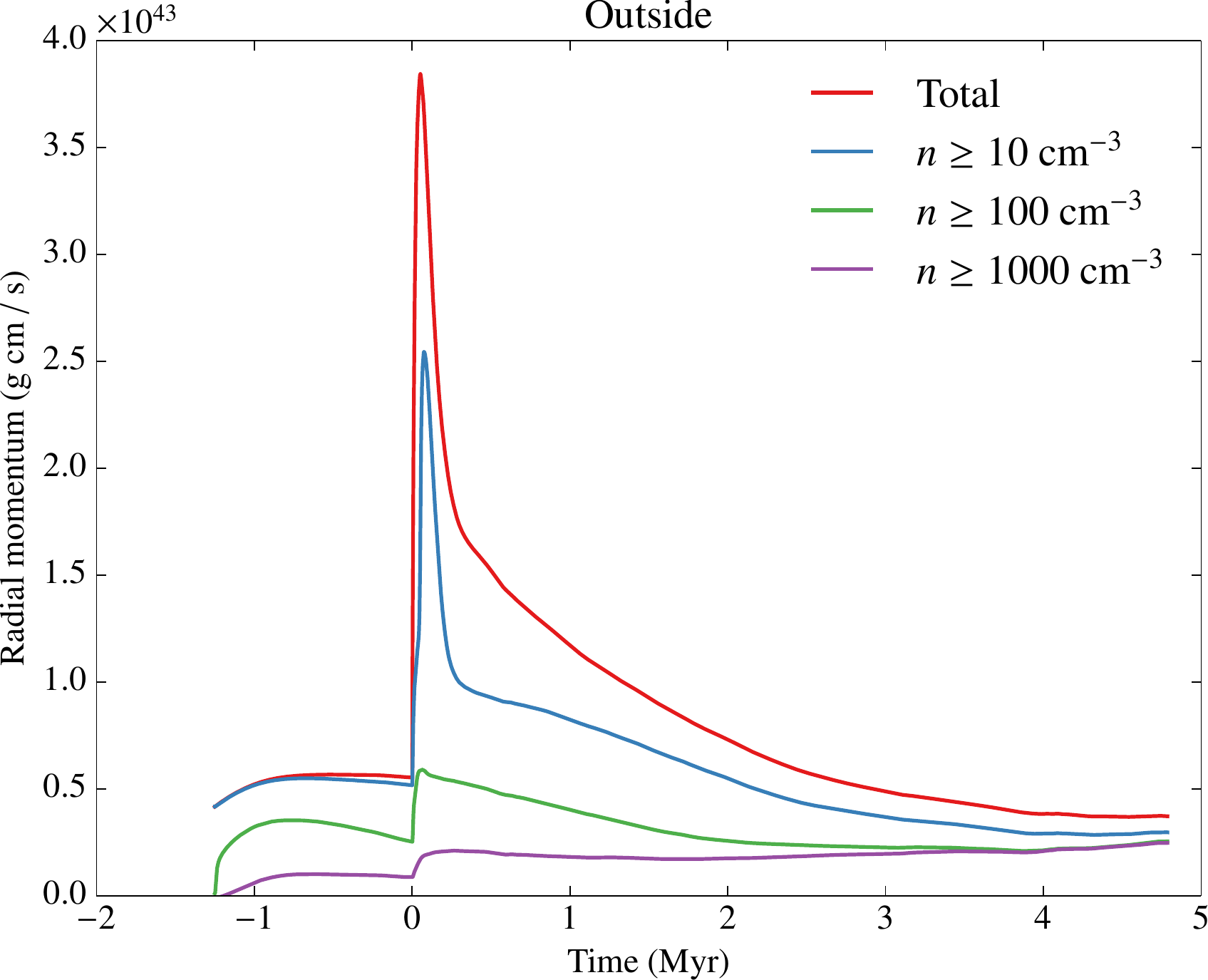}\\
            \includegraphics[width=75mm]{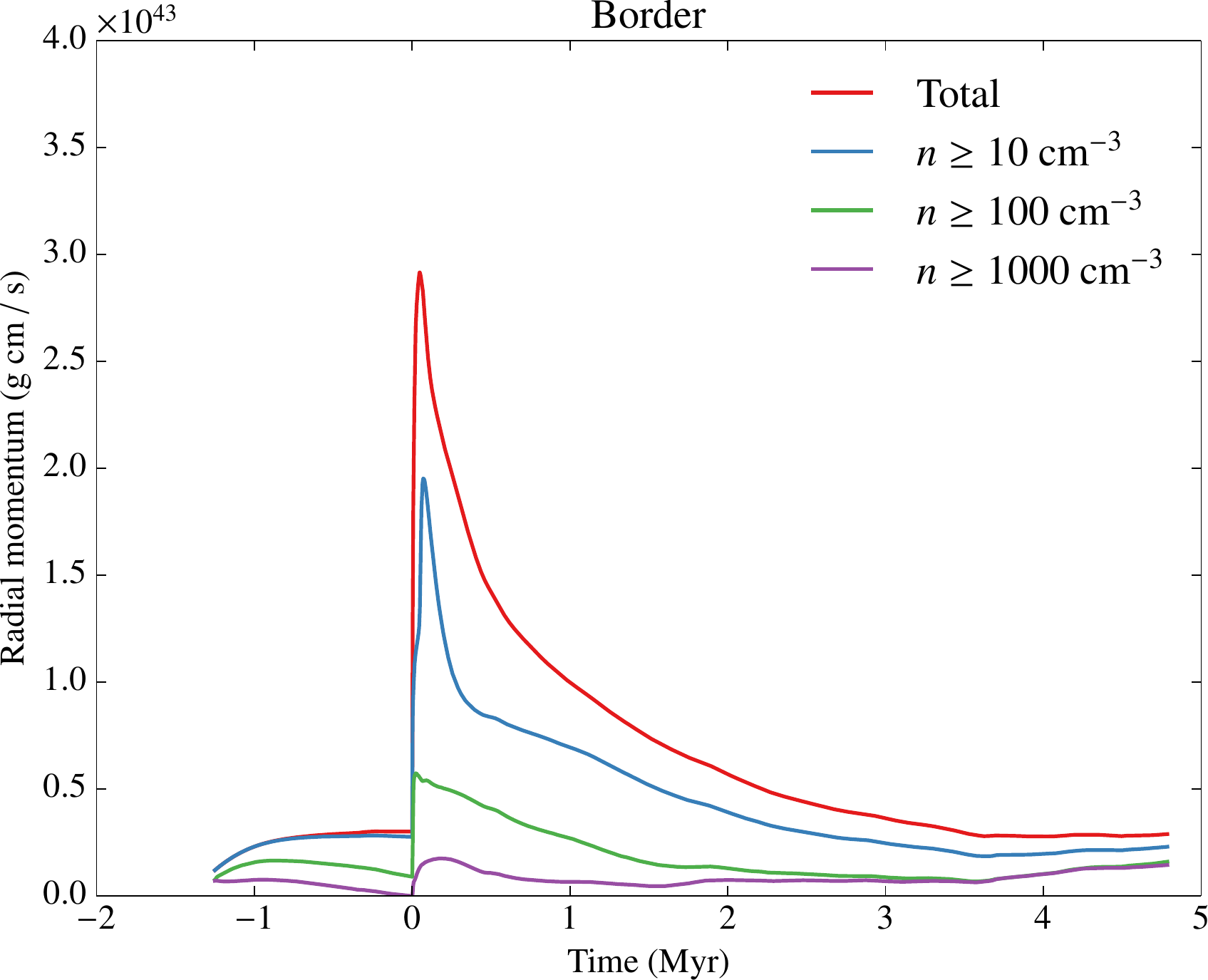}\\
            \includegraphics[width=75mm]{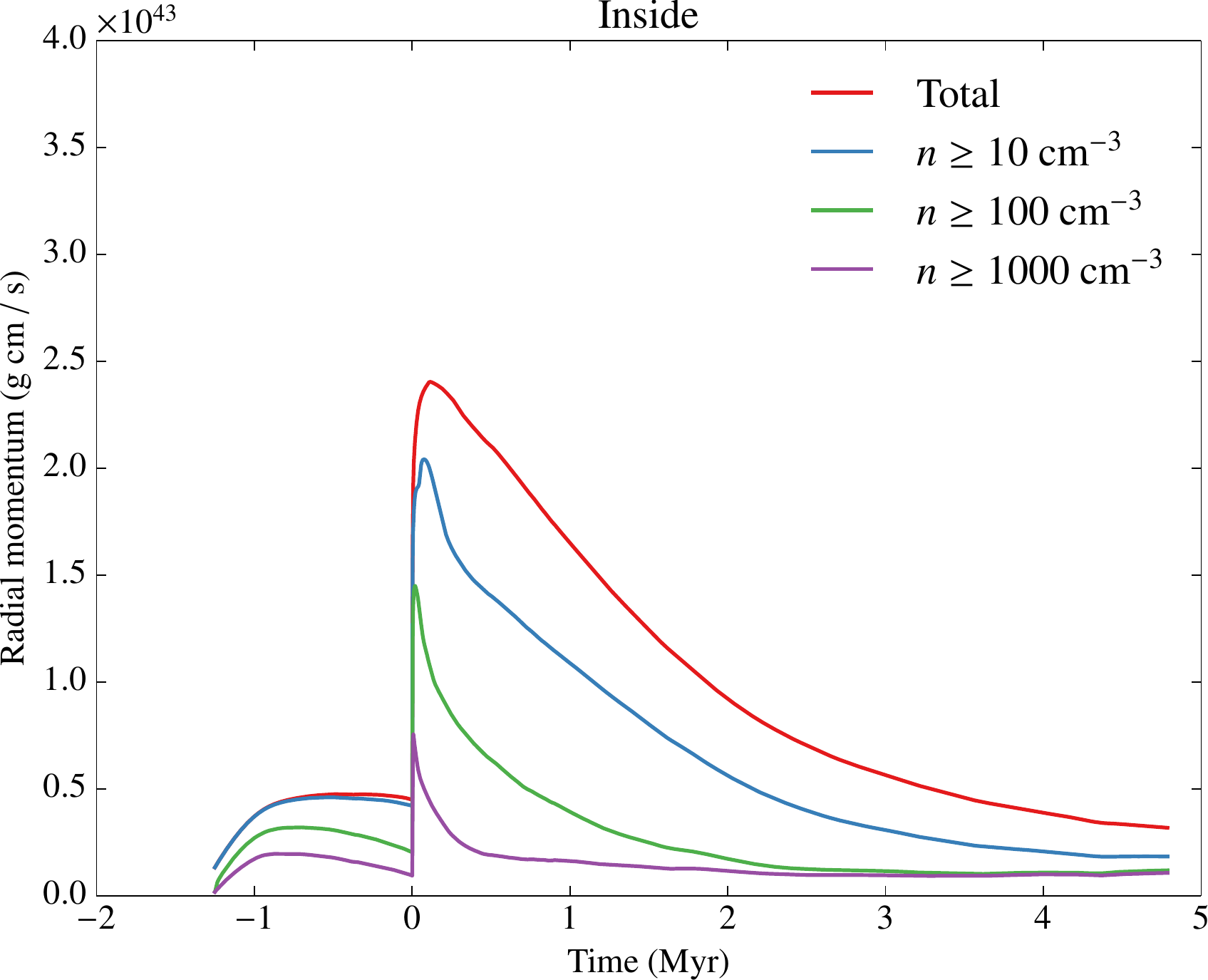}
         \end{center}
         \caption{Evolution of momentum for densities above thresholds $10,
            100,$ and $1000\ \mathrm{cm}^{-3}$ in the outside, border, and
            inside cases.}
         \label{graph-mom-thr}
      \end{figure}

      Figure~\ref{graph-mom-thr} shows the injected radial momentum for three
      density thresholds and the three supernova locations. Significantly less
      momentum is injected in denser regions. For the inside runs and for the
      $10^2$ and $10^3\ \mathrm{cm^{-3}}$ density thresholds, the amount of
      momentum is about 1.2 and $0.7 \times 10^{43}\ \mathrm{g\ cm\ s^{-1}}$
      respectively; these values are about $0.3$ and $0.1 \times 10^{43}\
      \mathrm{g\ cm\ s^{-1}}$ for the outside runs.

      In the case when the supernova explosion takes place inside the cloud,
      momentum is transferred to high-density medium more efficiently than in
      the other cases, probably because the supernova remnant is trapped within
      the dense gas of the cloud, whereas in the other cases a significant part
      of the remnant moves through diffuse medium. The momentum associated with
      dense gas quickly drops because this dense high momentum gas quickly
      reexpands and therefore becomes diffuse, in good agreement with the bottom
      panel of Fig.~\ref{graph-mass-thr}.

      This shows once again that the impact supernovae have on the cloud
      significantly depends on their positions. While the differences between
      the inside and outside runs show that a shift of a few pc in distance can
      make a significant difference to the impact a supernova can have on the
      clouds, supernovae exploding even farther away from the clouds would
      obviously have an even weaker impact on the dense gas. This highlights the
      fact that while the amount of momentum injected into the ISM is a very
      useful piece of information, it is nevertheless vastly insufficient. A
      more detailed knowledge of exactly where a supernova occurs is necessary
      in order to quantify the real influence it has on the ISM.

%__________________________________________________________________

    \subsection{An analytical estimate}
    \label{sect-mass-ej}

       To estimate the fraction of mass, $f_m$ that is expelled by the supernova
       explosion, we can simply compare the amount of momentum that is delivered
       by the supernova and the momentum necessary to unbind a mass $f_m M_c$
       from the cloud of mass $M_c$ and radius $R_c$. The escape velocity is
       given by
       \begin{equation}
          v_{esc} = \sqrt {2 M_c G \over R_c},
       \end{equation}
       therefore we must have
       \begin{gather}
          {p \over f_m M_c} \simeq v_{esc}, \\
          f_m = {p R_c^{1/2} \over \sqrt{2} M_c^{3/2} G^{1/2}}.
       \end{gather}
       Taking $M_c \simeq 10^4 \, M_\odot$, $p \simeq 10^{43}\ \mathrm{g\ cm\
       s^{-1}}$, and $R_c \simeq 5\ \mathrm{pc}$ (estimated from
       Fig.~\ref{rt-turb}), we get
       \begin{eqnarray}
          f_m \simeq 1.2 {p \over 10^{43} \mathrm{g\ cm\ s^{-1}}}
          \left( { R_c \over 5\ \mathrm{pc}} \right)^{1/2} \left( {M_c \over
          10^4\ \mathrm{M_\odot}} \right) ^{-3/2},
       \end{eqnarray}
       which for a cloud of $1.25 \times 10^4\ \mathrm{M_\odot}$ (as estimated
       from Fig.~\ref{graph-mass-thr}, bottom panel) gives $f_m = 0.87$.

       This value is in reasonable agreement with the numbers inferred from
       Fig.~\ref{graph-mass-thr}, since the mass of gas above $10^3\
       \mathrm{cm^{-3}}$ in density in the inside run at $5\ \mathrm{Myr}$ is
       about $6 \times 10^3\ \mathrm{M_\odot}$, while the momentum injected in
       this dense gas (from Fig.~\ref{graph-mom-thr} at $t = 0$) is $0.7 \times
       10^{43}\ \mathrm{g\ cm / s}$, leading to a fraction of expelled dense gas
       around $0.6$. As mentioned above in the case without feedback the mass of
       the dense gas is typically 2 times larger than in the inside run (bottom
       panel). Given the complexity of the problem it is difficult to look for a
       more quantitative prediction.

       Of course, this estimate is valid for a single supernova event while in
       practice their number should be roughly equal to the total mass of stars
       divided by about $120\ \mathrm{M_\odot}$. The momentum, $p$, must
       therefore be multiplied by the number of supernovae events $N_s$.  Their
       total number will eventually be on the order of $\simeq (1-f_m)*
       M_c/120$, leading to 
       \begin{eqnarray}
          f_m \simeq \frac{1}{1 +  10^{-2} \left({p \over 10^{43} \mathrm{g\ cm\ s^{-1}}} \right)^{-1}
             \left({ R_c \over 5\
             \mathrm{pc}} \right)^{-1/2} \left( {M_c \over 10^4\ \mathrm{M_\odot}}
             \right) ^{1/2} },
       \end{eqnarray} 
       where $p$ is the momentum that is effectively injected in the star
       forming gas.  As discussed above, this value may vary substantially from
       one supernova to another.  This value of $f_m$ is obviously an upper
       estimate as it would lead to very low star formation efficiency, $1-f_m$.
       In practice, since supernovae typically arrive after $10$ -- $20\
       \mathrm{Myr}$, a significant amount of mass has already been converted
       into stars. It is clear, however, that if the supernovae are still
       sufficiently embedded into the molecular cloud when the massive stars
       start exploding, only a few of them will be enough to disperse the cloud
       entirely. As discussed above the efficiency depends on the correlation
       between the massive stars and the dense gas. As emphasized in other works
       \citep{matzner2002,dale+2013}, ionising radiation may have already pushed
       away the surrounding gas. It must be kept in mind, however, that ionising
       radiation has little impact on dense material. Therefore, detailed
       investigations are required to conclude.

       Finally, we note that in their paper, \citet{rogers2013} report the mass
       flux at their box boundaries. Clearly, the supernova significantly
       dominates the effect of the stellar winds (see their Fig. 10). The
       integrated  mass flux is a few $10^3$ M$_\odot$ and therefore comparable
       to the cloud mass. Since they also include losses from the red supergiant
       and Wolf-Rayet phase, it is hard to infer the amount of mass lost because
       of the supernova in this work, but the values are similar to our results. 

%__________________________________________________________________

\section{Conclusions}\label{sect-concl}

   We have performed a series of numerical simulations to study supernova
   explosions in the interstellar medium. We considered both uniform density
   medium and turbulent star forming clouds and ran hydrodynamical and MHD
   simulations.

   In good agreement with previous works and simple analytical considerations,
   we found that the total amount of momentum that is delivered in the
   surrounding ISM is weakly dependent on the density. This is true both for a
   uniform density medium and for a turbulent cloud.

   However, the impact a supernova has on a molecular cloud significantly
   depends on its location. If it is located outside the molecular cloud, its
   impact on the dense gas remains fairly limited and only a small percent of
   the momentum is given to the dense gas. When the supernova explodes inside
   the molecular cloud, up to one half of the momentum can be given to the dense
   gas. Consequently, supernovae can reduce significantly the mass of the cloud
   when they explode inside while they will barely affect the amount of dense
   gas if they explode outside. For the conditions we explore, that is to say a
   $10^4\ \mathrm{M_\odot}$ molecular cloud, we find that up to half of the mass
   can be removed by one supernova explosion. Simple analytical considerations
   suggest that these results can be understood by comparing the amount of
   momentum delivered with the momentum required to escape the gravitational
   potential of the cloud.

   The magnetic field has an overall weak impact on the mutual influence between
   molecular clouds and supernovae. In particular, it does not influence the
   total amount of momentum delivered onto the ISM. It tends, however, to
   enhance the effect a supernova has on the cloud when it is located inside.
   For the conditions we explore, the momentum injected in the dense gas
   increases by about $50\,\%$ and the mass that is removed is in approximately
   the same proportion.

   Our results suggest that the influence that supernovae have on molecular
   clouds and in particular their ability to regulate the star formation in
   galaxies, depends on their exact location.

%__________________________________________________________________

\begin{acknowledgements}
   It is a pleasure to thank Eve Ostriker for stimulating discussions and
   insightful suggestions, Anne Decourchelle and Jean Ballet for enlightening
   elements regarding the observations of supernovae remnants and Matthieu
   Gounelle for triggering our interest in supernova explosions.  We thank the
   anonymous referee for a careful reading of the manuscript which has
   significantly improved the paper. This work was granted access to HPC
   resources of CINES under the allocation x2014047023 made by GENCI (Grand
   Equipement National de Calcul Intensif). PH acknowledges the financial
   support of the Agence Nationale pour la Recherche through the COSMIS project.
   This research has received funding from the European Research Council under
   the European Community's Seventh Framework Programme (FP7/2007-2013 Grant
   Agreement no. 306483 and no. 291294).
\end{acknowledgements}

% for the bibliography
\bibliography{refs}{}
\bibliographystyle{aa} % style aa.bst

\Online

\begin{appendix}
   \section{Kinetic energy injection}
      \label{sect-kin}

\begin{figure}
         \begin{center}
            \includegraphics[width=80mm]{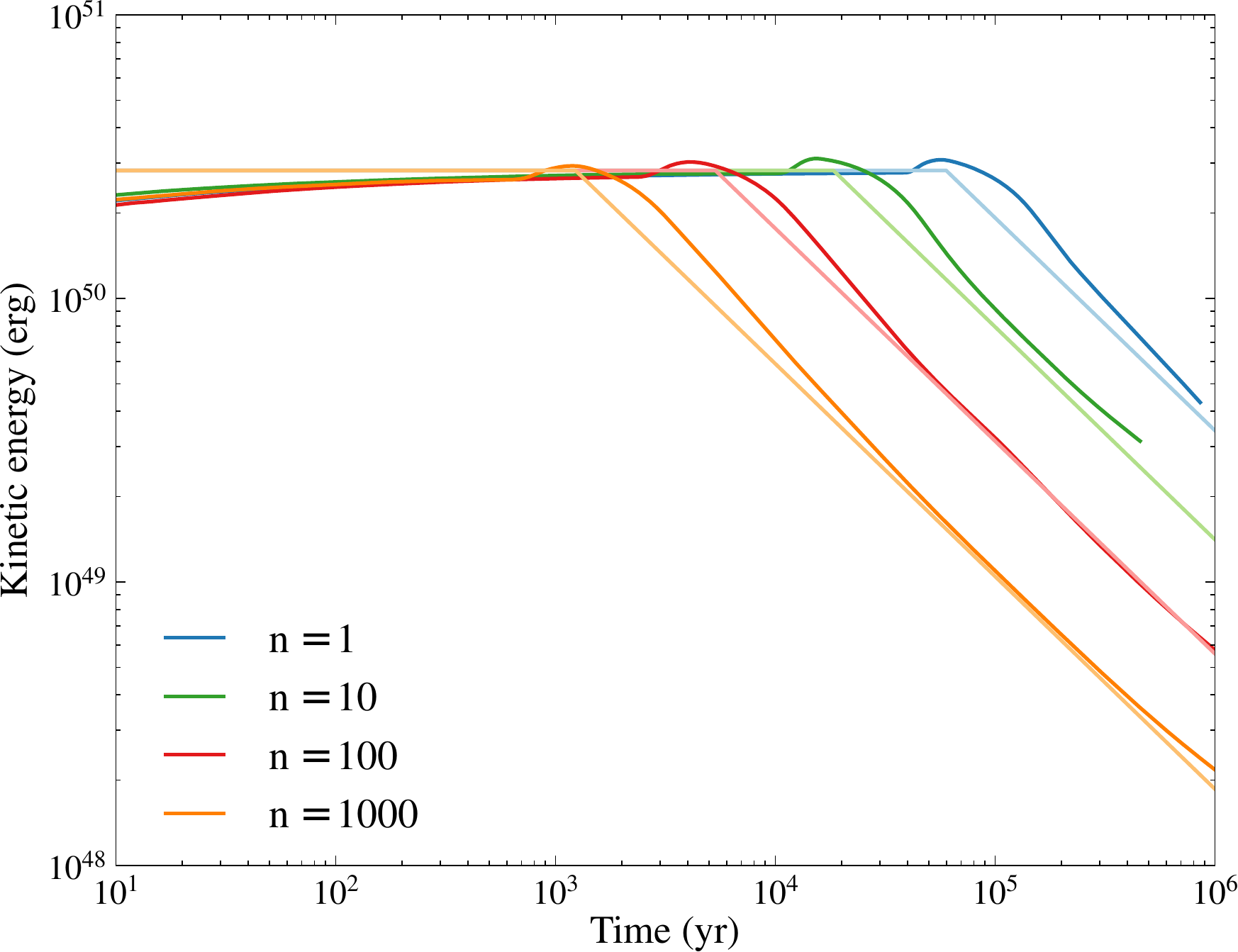}
         \end{center}
         \caption{Total kinetic energy for the uniform simulations.
            The thin straight lines correspond to the analytical trends
            described in Sect.~\ref{analyt}.
         }
         \label{graph-kinetic-uniform}
      \end{figure}

      It is also worth studying the amount of kinetic energy that is injected
      into the ISM during the supernova explosions. Figure
      \ref{graph-kinetic-uniform} shows the total kinetic energy as a function
      of time. As with the momentum plots described before, we see the adiabatic
      phase where energy is conserved (the kinetic energy being a constant
      fraction of the total energy in this phase), the shell formation, and the
      snowplow phase where the energy of the hot dense shell is radiated away,
      approximately following the momentum-conserving snowplow model $E_K
      \propto t^{-3/4}$. The ratio between total and kinetic energy is about
      $0.2$ -- $0.3$ in the adiabatic phase in good agreement with Sedov's
      model.

      For the momentum-conserving snowplow model, the shell radius evolves with
      time as $R_s \propto t^{1/4}$. This stems from the fact that $p \propto
      R^3 v$, while $v=dR/dt$ and $p$ is nearly constant.  Therefore, the
      kinetic energy can be approximated by
      \begin{equation}
         E_{K,51} = 0.28 E_{51} \left(\frac{t_4}{2t_{tr,4}}\right)^{-3/4},
      \end{equation}
      where $E_{K,51}$ is the integrated kinetic energy in $10^{51}\
      \mathrm{erg}$, $E_{51}$ is the initial supernova energy in $10^{51}\
      \mathrm{erg}$, $t_4$ is the age of the remnant in $10^4\ \mathrm{yr}$, and
      $t_{tr,4}$ is the transition time in $10^4\ \mathrm{yr}$.

\begin{figure}
         \begin{center}
            \includegraphics[width=80mm]{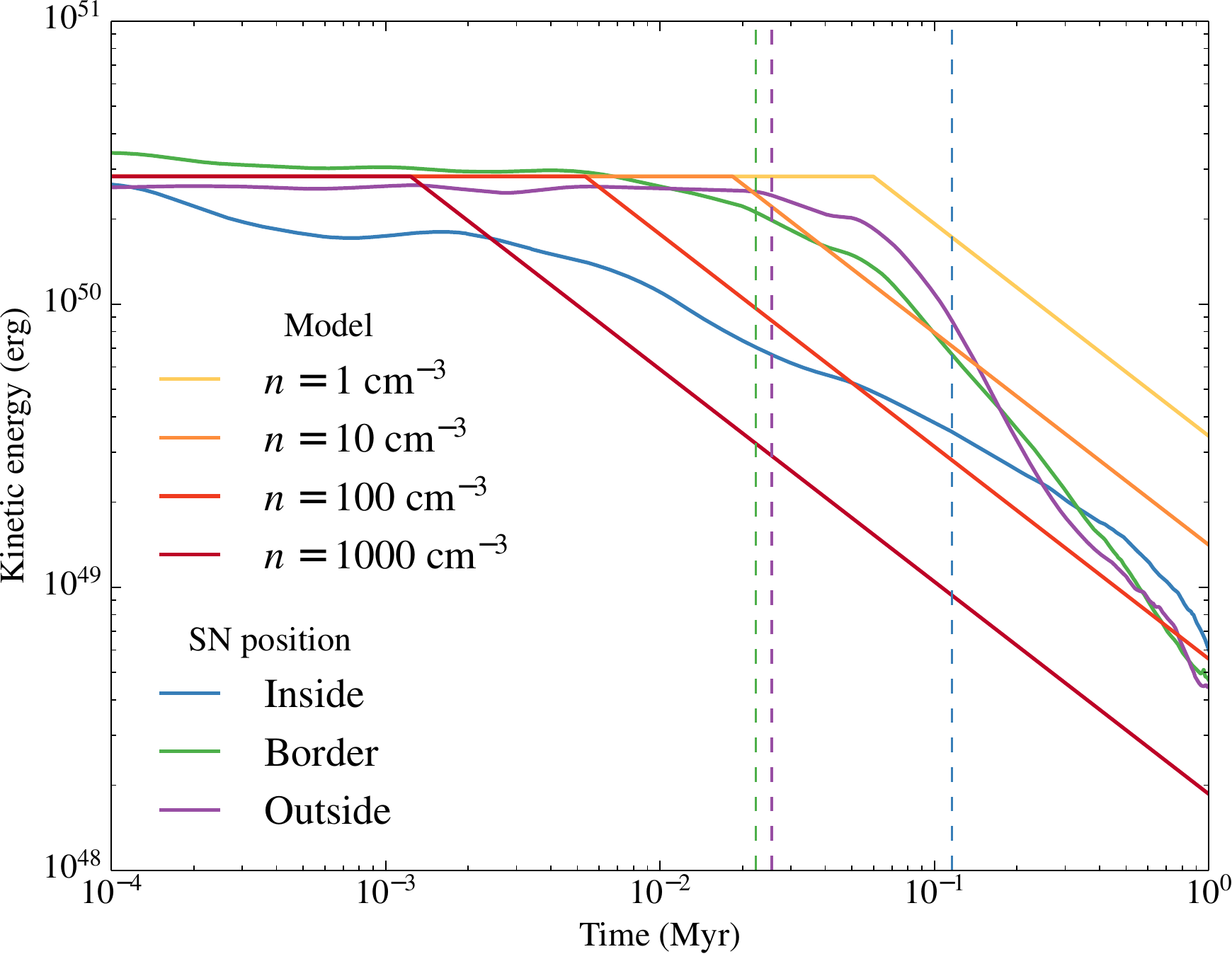}
         \end{center}
         \caption{Kinetic energy injection: comparison between turbulent
            simulations and our model. The vertical lines correspond to the time at
            which the gas starts leaving the computational domain (from left to right:
            border, outside, inside)}
         \label{graph-kinetic-turb}
      \end{figure}

      Figure \ref{graph-kinetic-turb} shows the evolution of kinetic energy in
      the turbulent case, compared to our simple model. The injected kinetic
      energy roughly corresponds to the uniform case in the Sedov-Taylor phase,
      and stays in the same order of magnitude in the radiative phase.

%      According to figure \ref{graph-kinetic-turb-raw}, the energy injection
%      efficiency, which we define as the fraction between the injected energy just
%      before the cloud collapse and the supernova initial energy, is of the
%      order of one percent.

%      \begin{figure}
%         \begin{center}
%            \includegraphics[width=80mm]{pic/turb/rkin_hyd.pdf}
%         \end{center}
%         \caption{Total kinetic energy evolution in the turbulent simulations.}
%         \label{graph-kinetic-turb-raw}
%      \end{figure}

%__________________________________________________________________

   \section{Density distributions}
      \label{sect-pdf}

\begin{figure}
         \begin{center}
            \includegraphics[width=80mm]{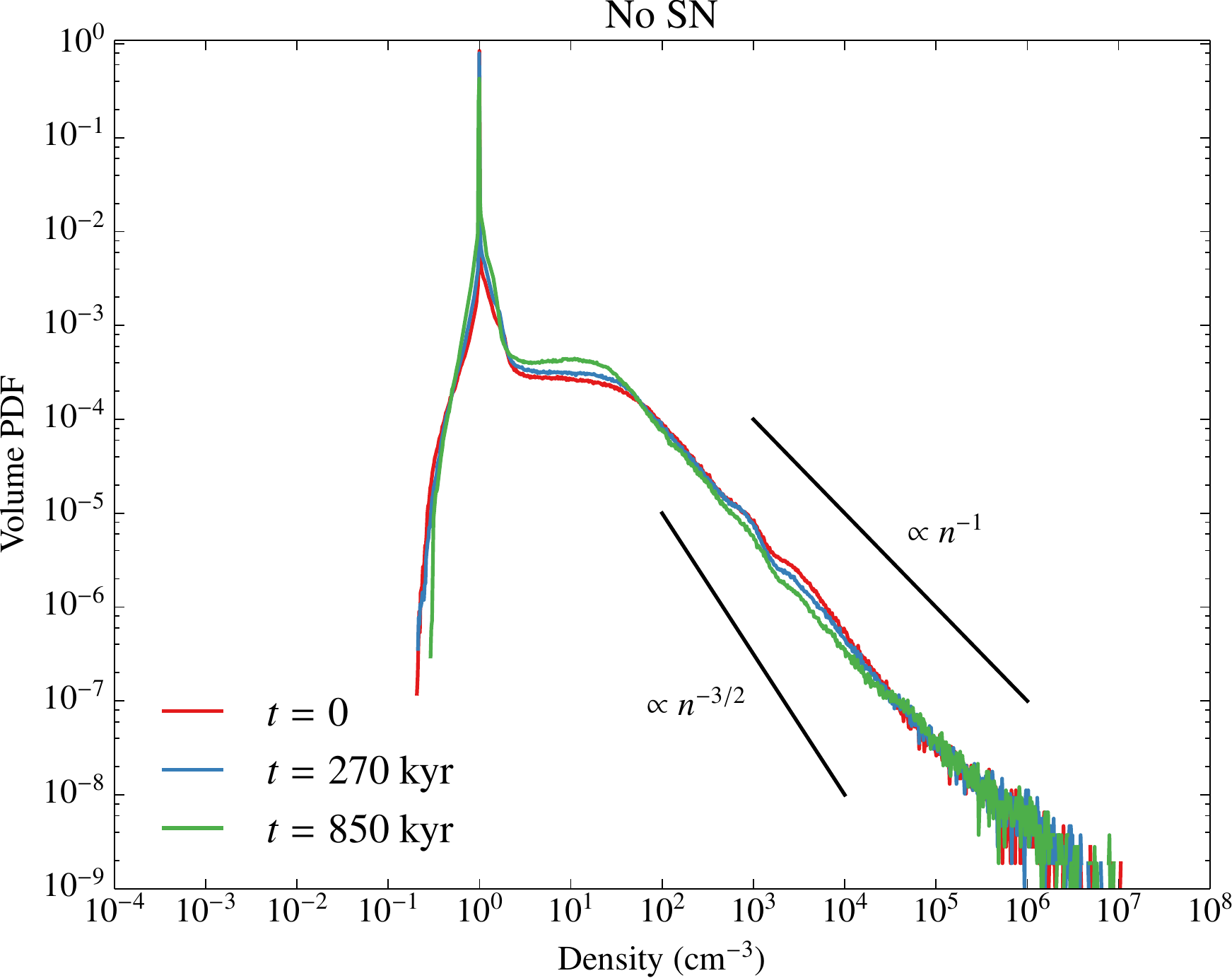}\\
            \includegraphics[width=80mm]{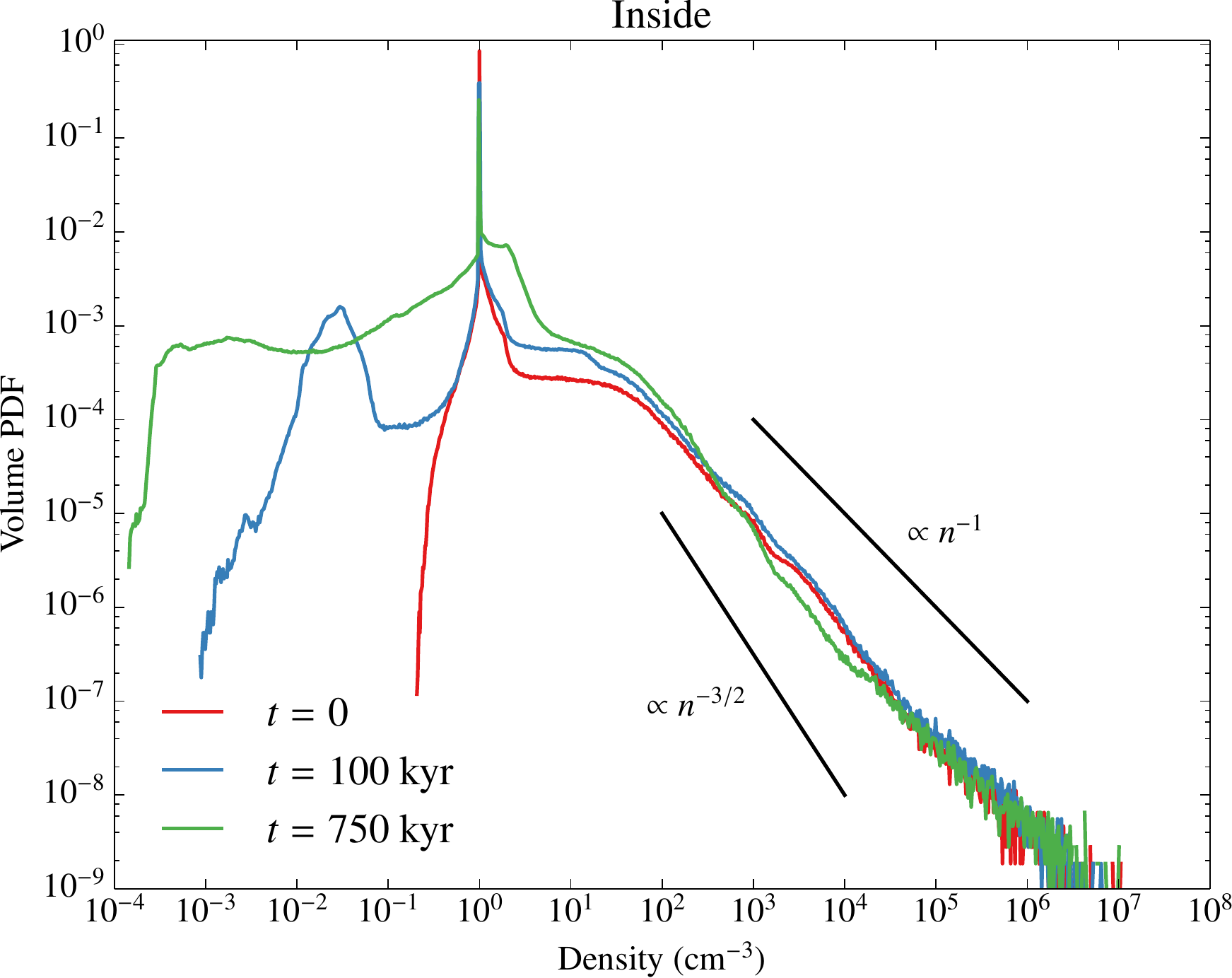}
         \end{center}
         \caption{Density probability distribution just before the explosion and
            at two later times. Top panel displays the case without supernova
            and bottom panel shows the inside run. While significant differences
            are seen in the diffuse gas distribution, the high-density tail is
            largely unchanged by the supernova explosions.}
         \label{graph-pdf}
      \end{figure}

      Figure~\ref{graph-pdf} shows the density probability distribution
      functions in the runs without supernova (top panel) and with a supernova
      inside (bottom panel). As can be seen, a high-density power law with a
      slope between $-1$ and $-3/2$ develops. Such a power law has been found in
      simulations including gravity and turbulence \citep[e.g.,][]{kritsuk+2011}
      and is due to the collapse itself. The supernova does not change the
      high-density part of the distribution, but produces very diffuse gas and
      hot material. This is even more clearly seen in Fig.~\ref{graph-mass-thr},
      which shows the mass above various thresholds as a function of time in the
      four runs (without supernova, outside, border, and inside).

%_________________________________________________

   \section{Influence of the magnetic field}\label{sect-mag}

      \begin{figure}
         \begin{center}
            \includegraphics[width=80mm]{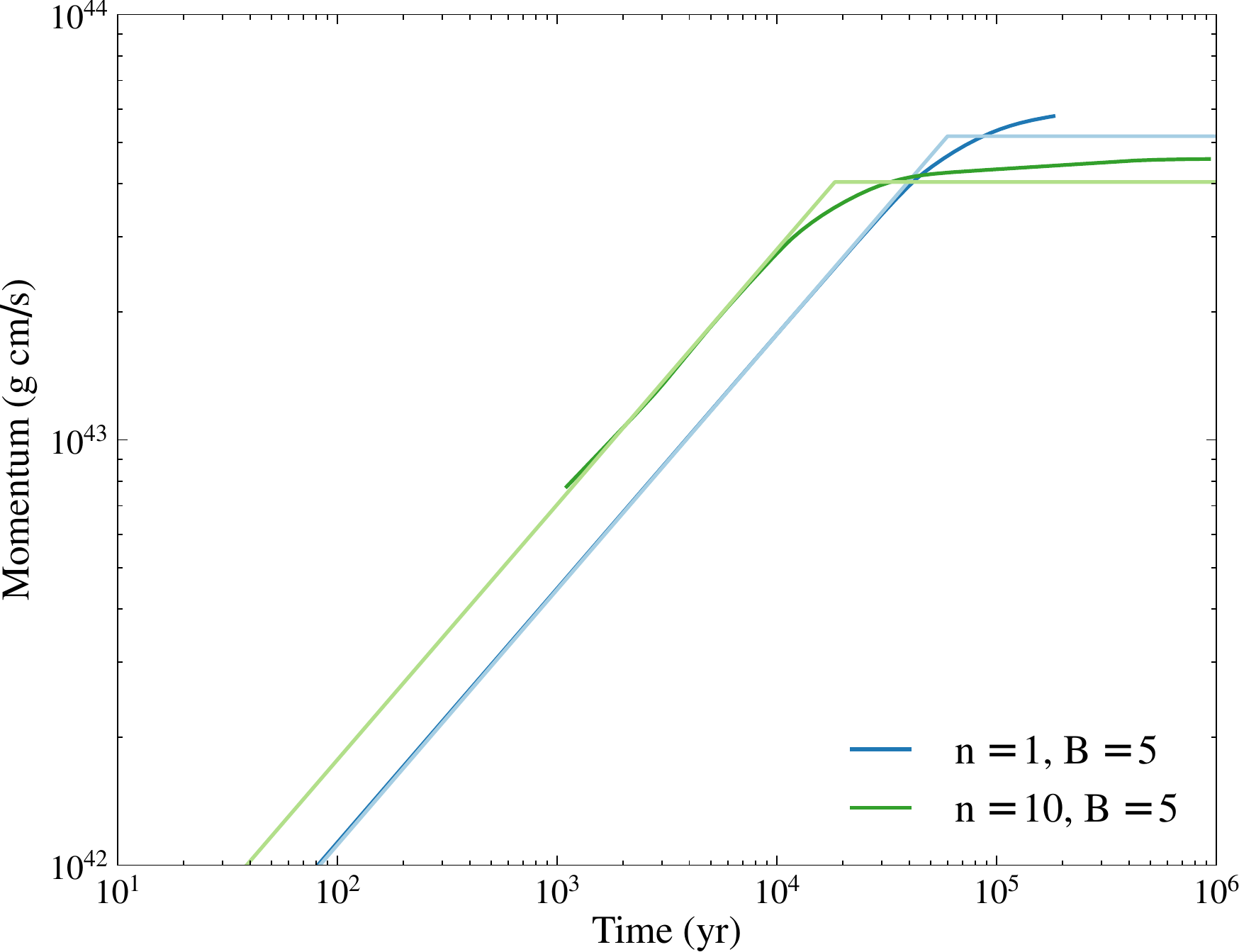}
         \end{center}
         \caption{Integrated radial momentum for the uniform MHD simulations.
            The thin straight lines correspond to the analytical trends
            described in Sect.~\ref{analyt}.
         }
         \label{graph-momentum-uniform-mhd}
      \end{figure}

      We now study the impact that a magnetic field can have on the supernova
      remnant influence on the ISM. We proceed as for the hydrodynamical case
      starting with the uniform configuration, and then investigate the
      turbulent case.

      \subsection{Uniform case}

         Figure~\ref{graph-momentum-uniform-mhd} shows the differences between
         momentum injection with an ambient uniform magnetic field (taken to be
         $5\ \mathrm{\mu G}$) and the model for $n = 1$ and $n = 10\
         \mathrm{cm^{-3}}$.  The values in the final stage are almost unchanged
         by the presence of magnetic field. This is relatively unsurprising
         since as seen from the turbulent simulations, the final momentum is
         relatively insensitive to density variations and complex geometry. In
         particular, the magnetic field does not alter very significantly the
         shock structure as long as it remains adiabatic, nor does it modify the
         cooling rate.

%         \begin{figure}
%            \begin{center}
%               \includegraphics[width=80mm]{pic/unif/kin_sn_unif_mhd_color.pdf}
%            \end{center}
%            \caption{Kinetic energy for the uniform MHD simulations.}
%            \label{graph-kinetic-uniform-mhd}
%         \end{figure}

%__________________________________________________________________

      \subsection{Turbulent case}

         To include a magnetic field in the turbulent cloud runs, we introduce a
         uniform field initially. Its intensity is about $5\ \mathrm{\mu G}$,
         which corresponds for the $10^4\ \mathrm{M_\odot}$ cloud to an initial
         mass-to-flux ratio of about 10. The initial magnetic field is, however,
         significantly amplified before the supernova is introduced.

         \begin{figure}
            \begin{center}
               \includegraphics[width=75mm]{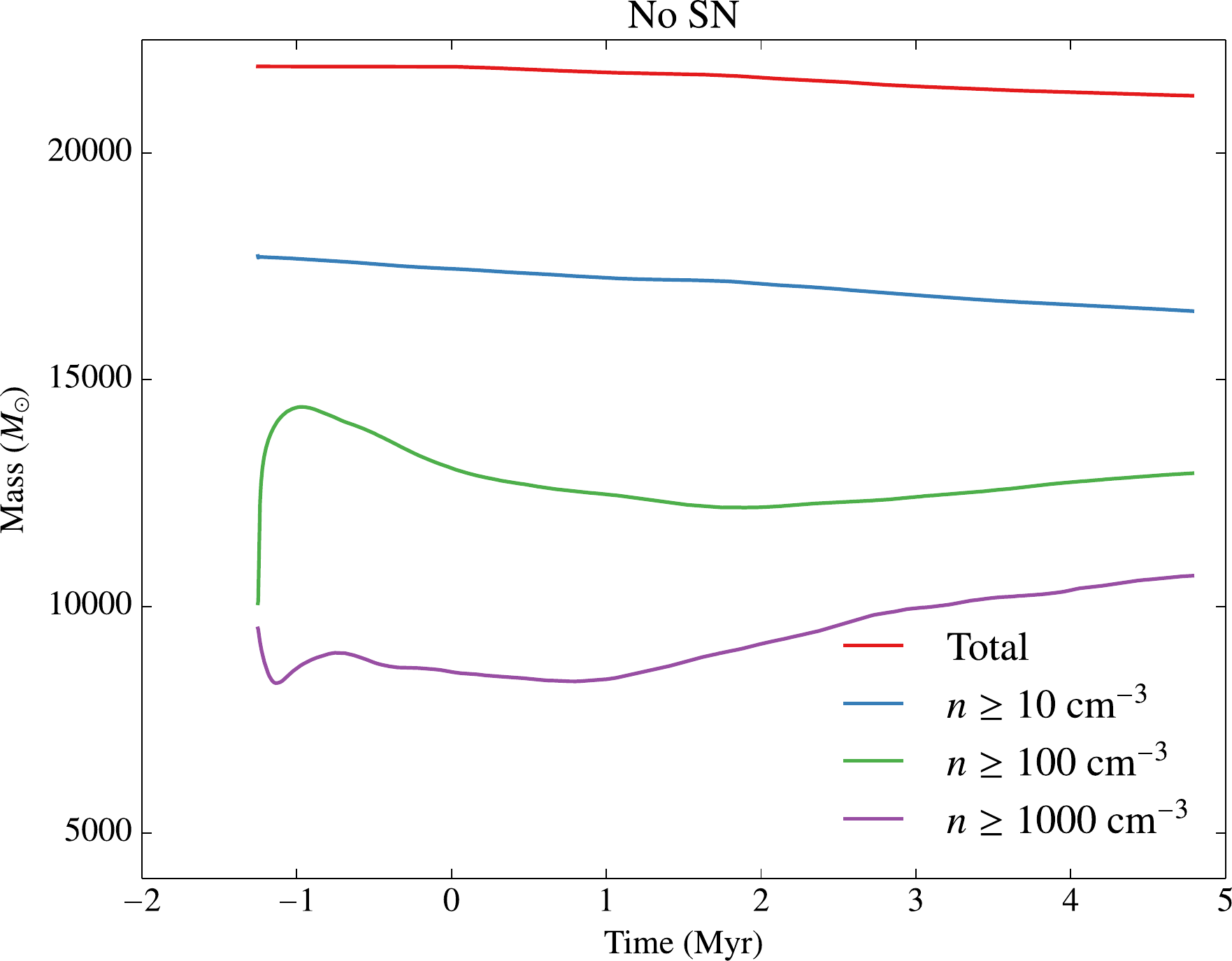}\\
               \includegraphics[width=75mm]{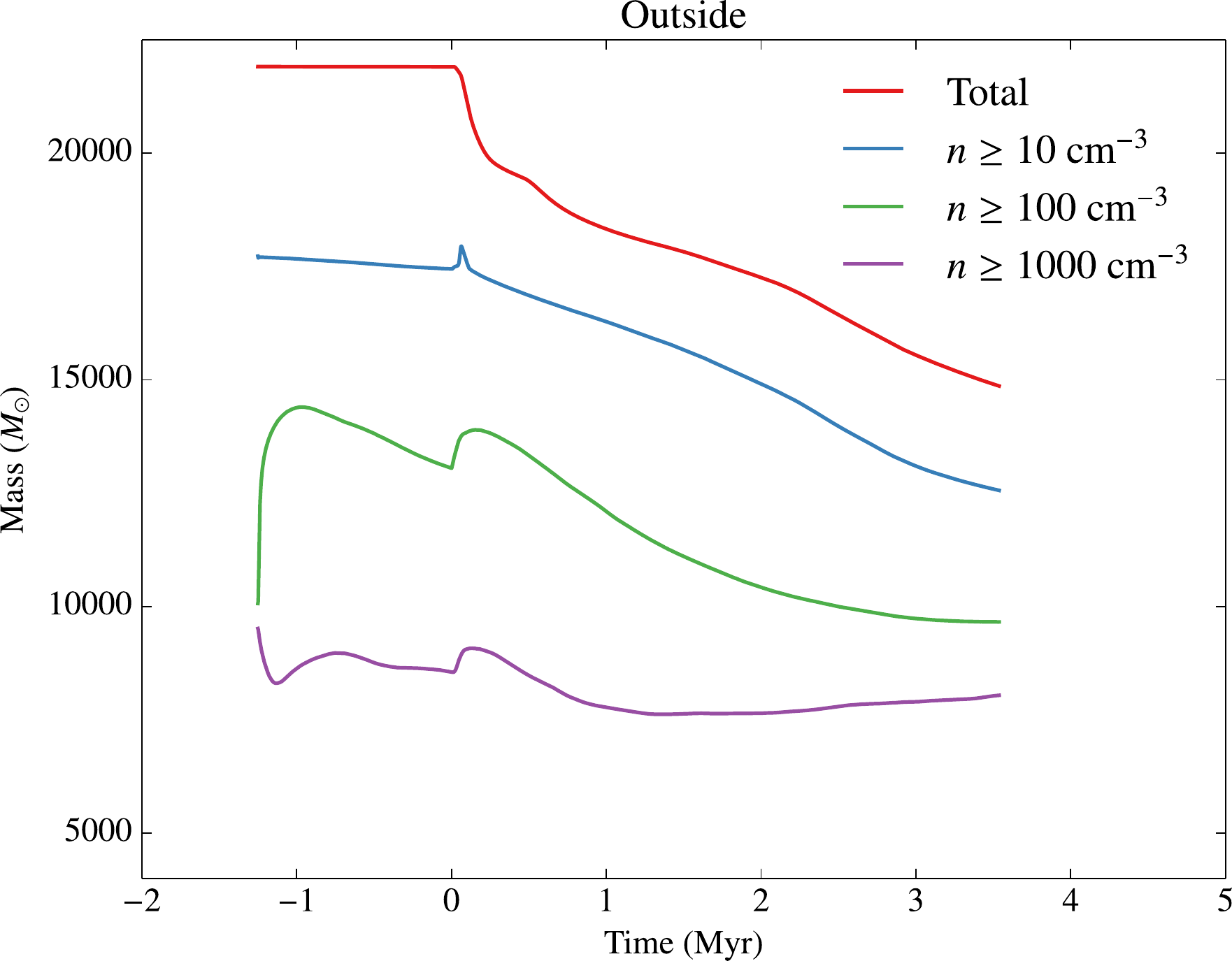}\\
               \includegraphics[width=75mm]{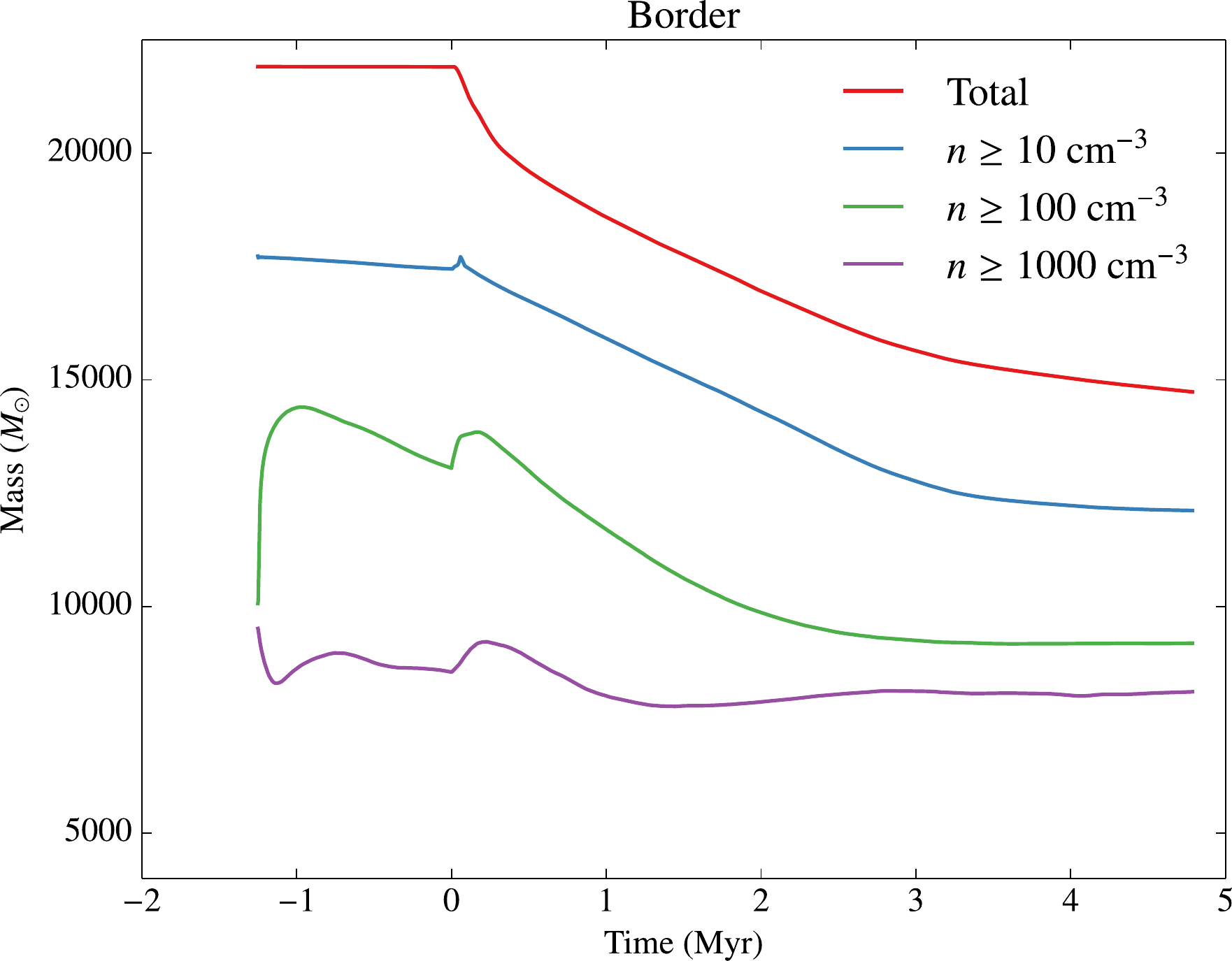}\\
               \includegraphics[width=75mm]{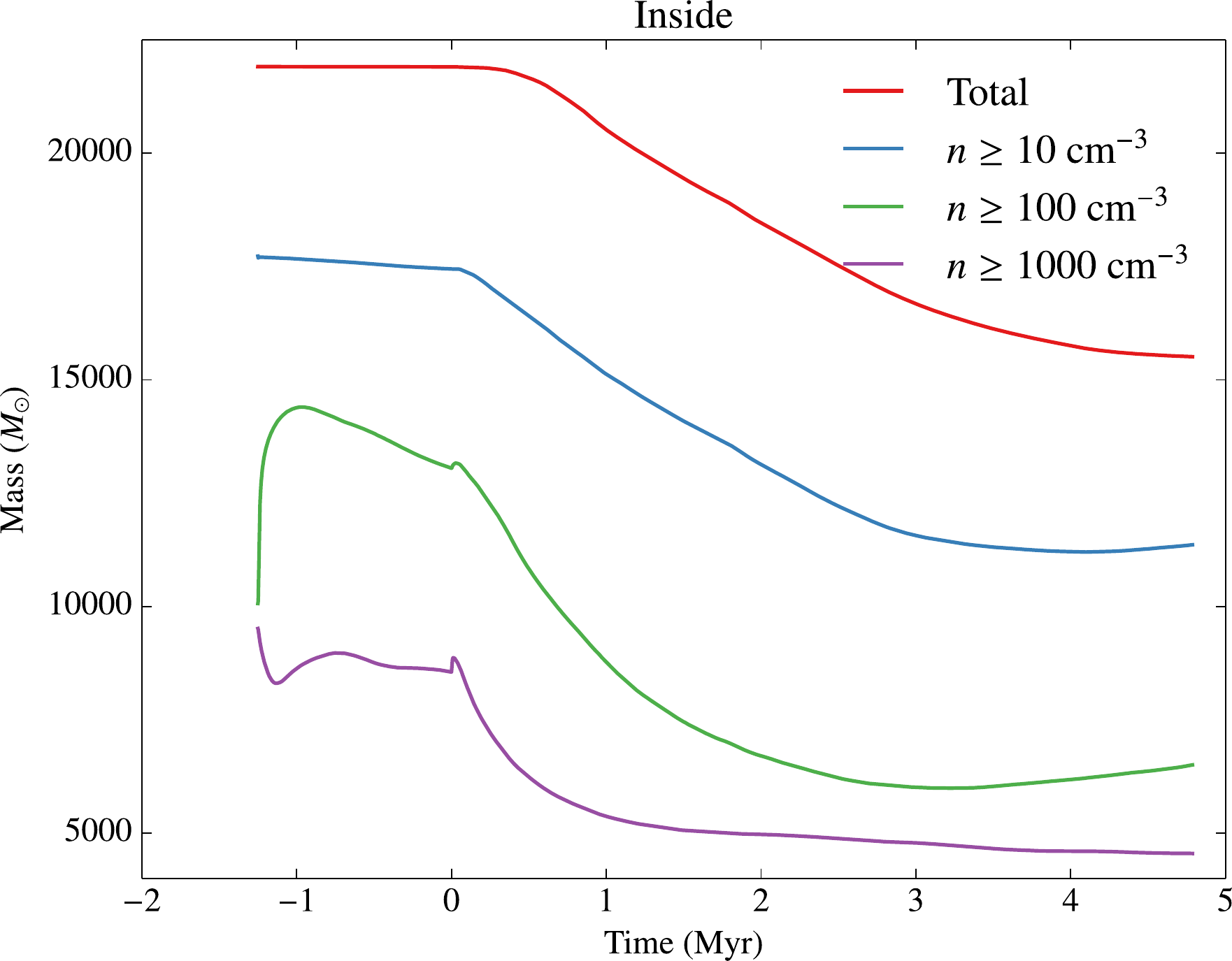}
            \end{center}
            \caption{MHD case. Mass above densities thresholds $10, 100, 1000\
               \mathrm{cm}^{-3}$ in the case without supernova (top panel) and
               outside (second panel), border (third panel) and
               inside (bottom panel) runs.}
            \label{graph-mass-mhd-thr}
         \end{figure}

         \begin{figure}
            \begin{center}
               \includegraphics[width=80mm]{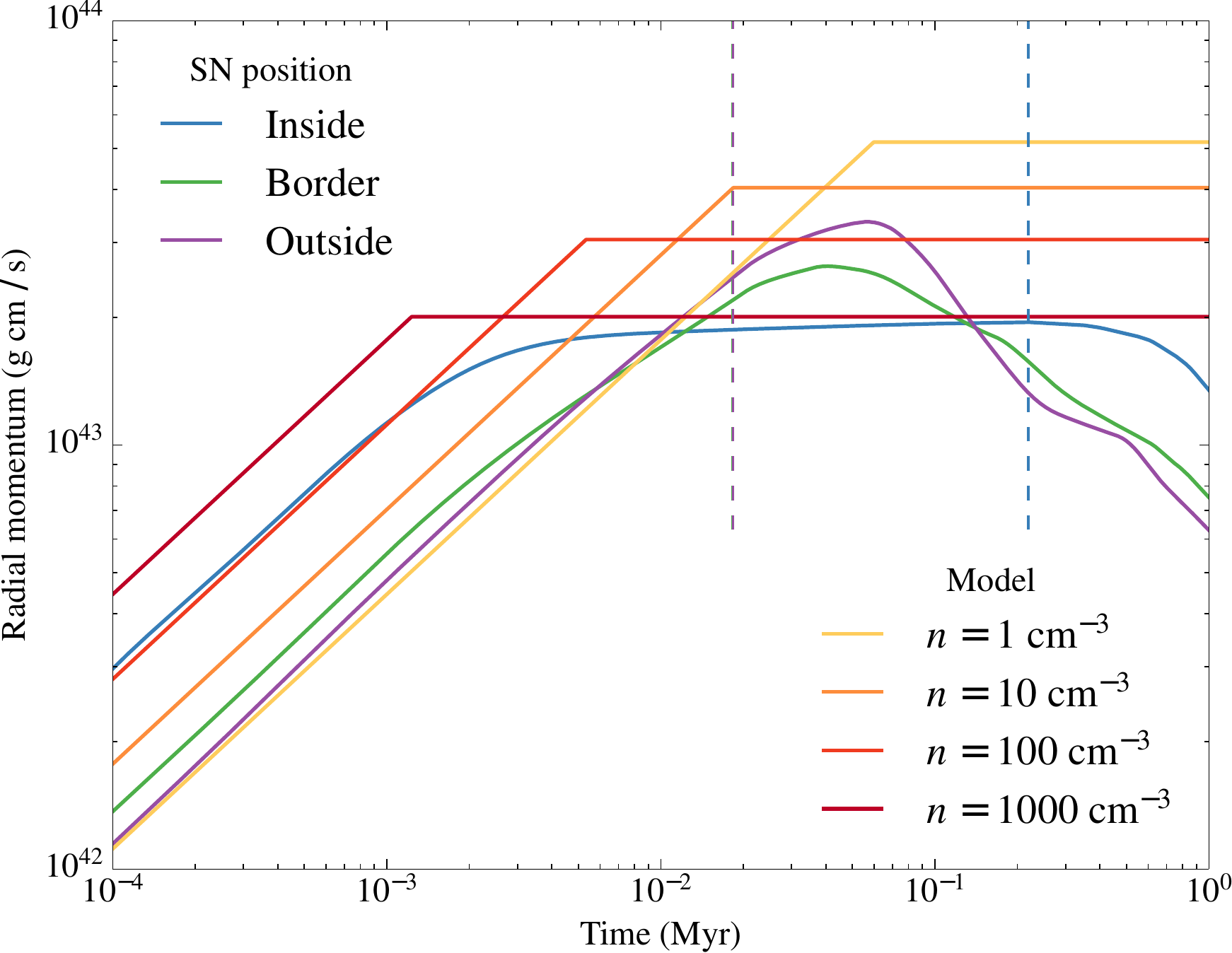}
            \end{center}
            \caption{Integrated radial injected momentum: comparison between
               turbulent MHD and our model. The vertical lines correspond to the first
               matter outflow for each case (from left to right: border, outside,
               inside)}
            \label{graph-momentum-turb-mhd}
         \end{figure}

         \subsubsection{Impact of the supernova remnant on the cloud}

            The evolution (not shown for conciseness) of the magnetized cloud is
            very similar to the hydrodynamical case. The supernova hot gas
            quickly escapes through the low-density material, whereas the
            high-density clumps are pushed away more slowly. A difference is
            that the propagation of the supernova in the diffuse medium is no
            longer spherical because it propagates more easily along the
            magnetic field lines as noted in earlier works
            \citep[e.g.,][]{tomisaka1998}.

%            \begin{figure*}
%               \begin{center}
%                  \includegraphics[width=160mm]{pic/turb/rt_mhd_0.pdf}\\
%                  \includegraphics[width=160mm]{pic/turb/rt_mhd_1.pdf}\\
%                  \includegraphics[width=160mm]{pic/turb/rt_mhd_3.pdf}
%               \end{center}
%               \caption{Column density maps for the turbulent MHD simulations. Top panel:
%                  no supernova, $0.25, 0.65, 1.0\ \mathrm{Myr}$ after injection time ;
%                  middle panel: supernova inside ; bottom panel: supernova outside,
%                  both after $100$, $200$ and $750\ \mathrm{kyr}$.}
%               \label{rt-turb-mhd}
%            \end{figure*}
 
%            \begin{figure*}
%               \begin{center}
%                  \includegraphics[width=160mm]{pic/turb/temp_mhd_0.pdf}\\
%                  \includegraphics[width=160mm]{pic/turb/temp_mhd_1.pdf}\\
%                  \includegraphics[width=160mm]{pic/turb/temp_mhd_3.pdf}
%               \end{center}
%               \caption{Temperature maps for the turbulent MHD simulations $1$, $20$ and
%                  $100\ \mathrm{kyr}$ after supernova injection time. Top panel: no
%                  supernova ; middle panel: supernova inside ; bottom panel: supernova
%                  outside (in this view the explosion takes place behind the cloud).}
%               \label{tmap-turb-mhd}
%            \end{figure*}

            Figure~\ref{graph-mass-mhd-thr} shows the mass above 3 density
            thresholds as a function of time for the four MHD runs. The run
            without supernova (top panel) is very similar to the corresponding
            hydrodynamical run (top panel of Fig.~\ref{graph-mass-thr}). In
            particular, the mass above $10^3\ \mathrm{cm^{-3}}$ is almost
            identical in the two runs.

            In the two MHD runs outside and border (second and third panels),
            the mass of gas above $10^3\ \mathrm{cm^{-3}}$ is smaller than in
            the hydrodynamical case by a factor of about $10$ -- $20\,\%$. The
            same is true for the inside run where it is seen that the mass above
            $10^3\ \mathrm{cm^{-3}}$ rapidly drops below $5000\
            \mathrm{cm^{-3}}$.

            This shows that the presence of a magnetic field tends to enhance
            the influence supernovae have on molecular clouds, probably because
            the magnetic field exerts a coupling between the different fluid
            particles within molecular clouds. Therefore, as some gas is pushed
            away by the high pressure supernova remnant, more gas is entrained.

%__________________________________________________________________

         \subsubsection{Momentum injection}

            Figure \ref{graph-momentum-turb-mhd} shows the evolution of the
            radial momentum (with respect to the supernova center) with time in
            the turbulent case with magnetic field. The evolution is very
            similar to the hydrodynamical case displayed in Fig.
            \ref{graph-momentum-turb} (top panel) and reasonably well described
            by the simple model presented in Sect.~\ref{analyt}. This confirms
            the result of the uniform density runs about the weak influence
            magnetic field has on the total momentum delivered to the ISM.

\begin{figure}
               \begin{center}
                  \includegraphics[width=75mm]{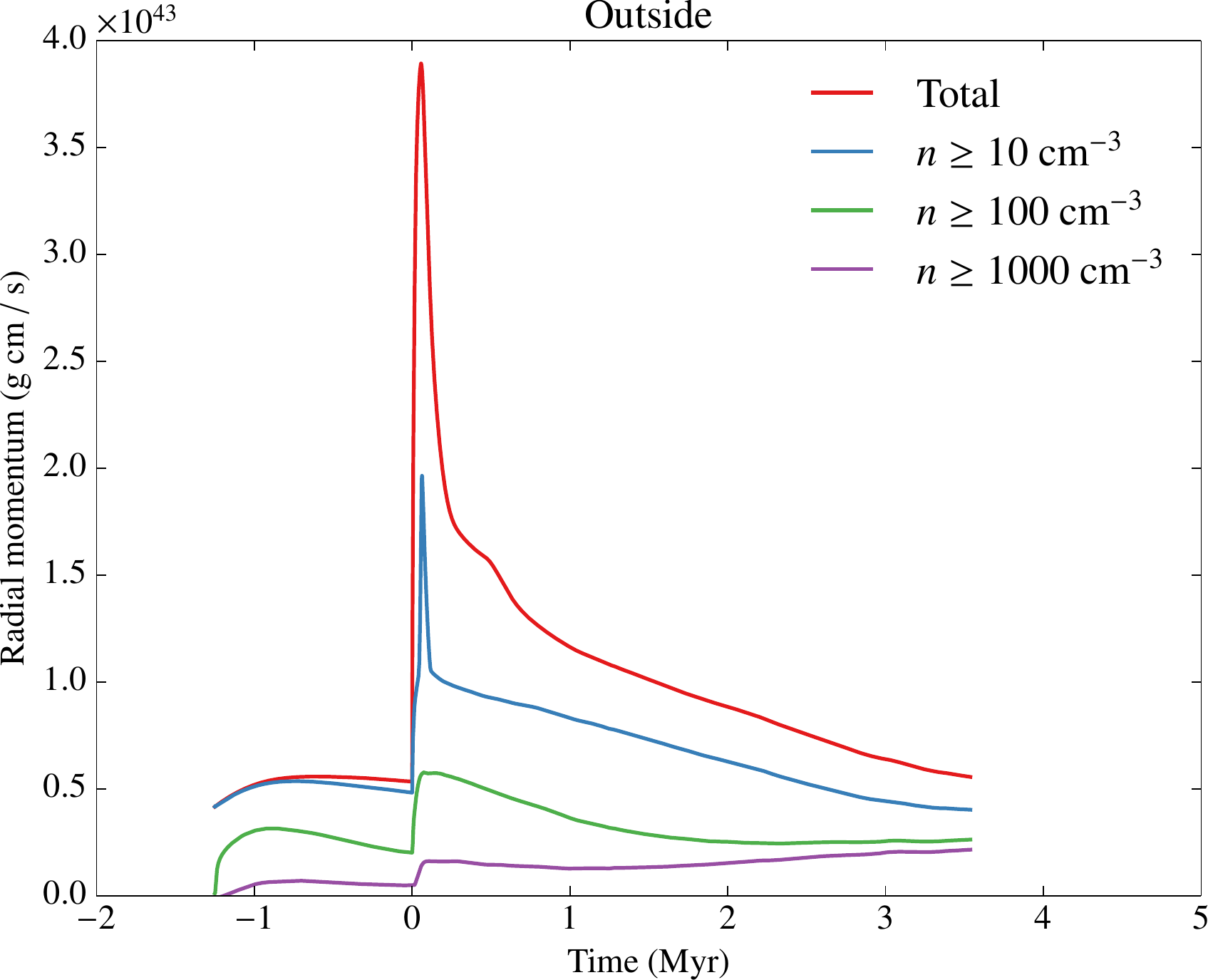}\\
                  \includegraphics[width=75mm]{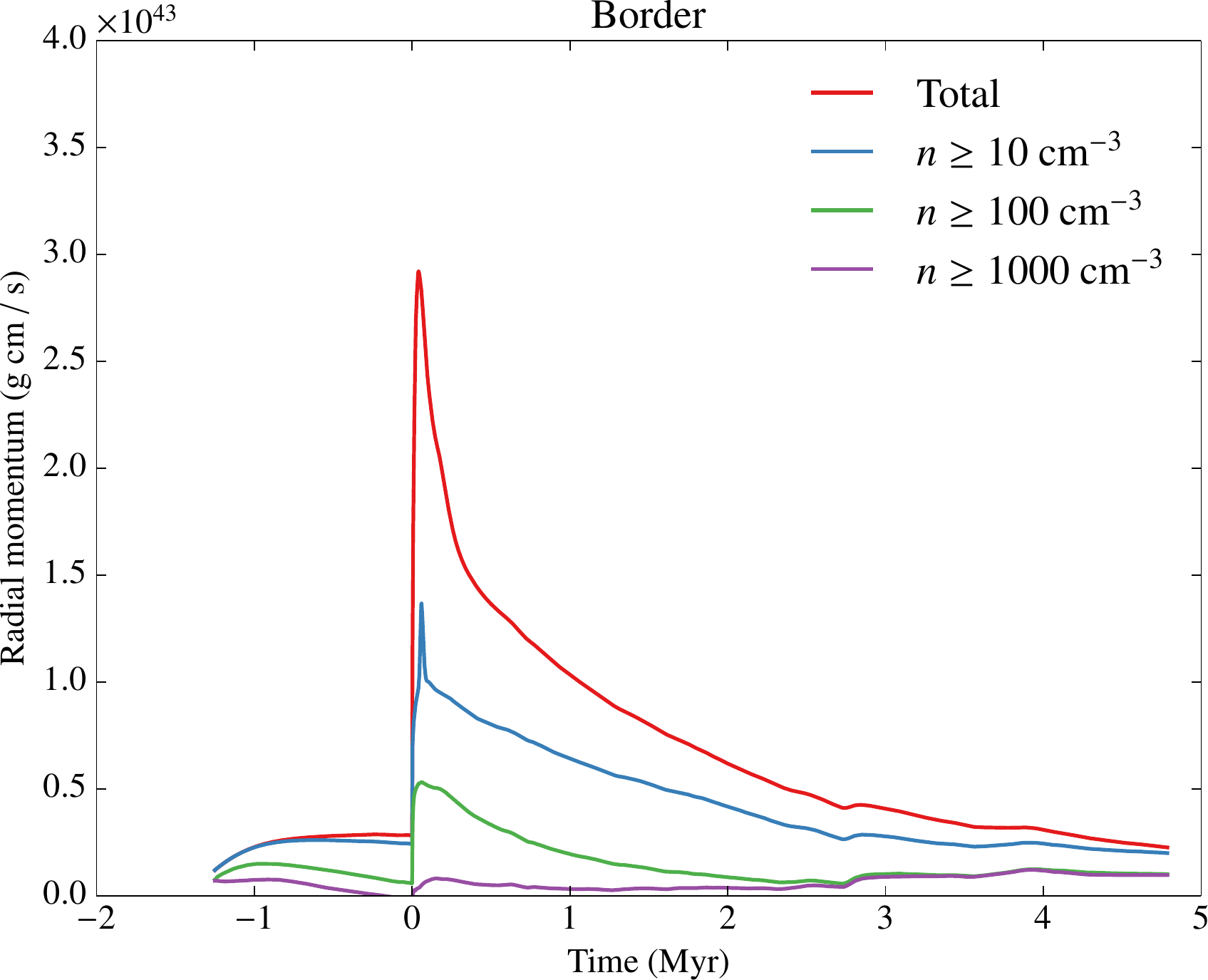}\\
                  \includegraphics[width=75mm]{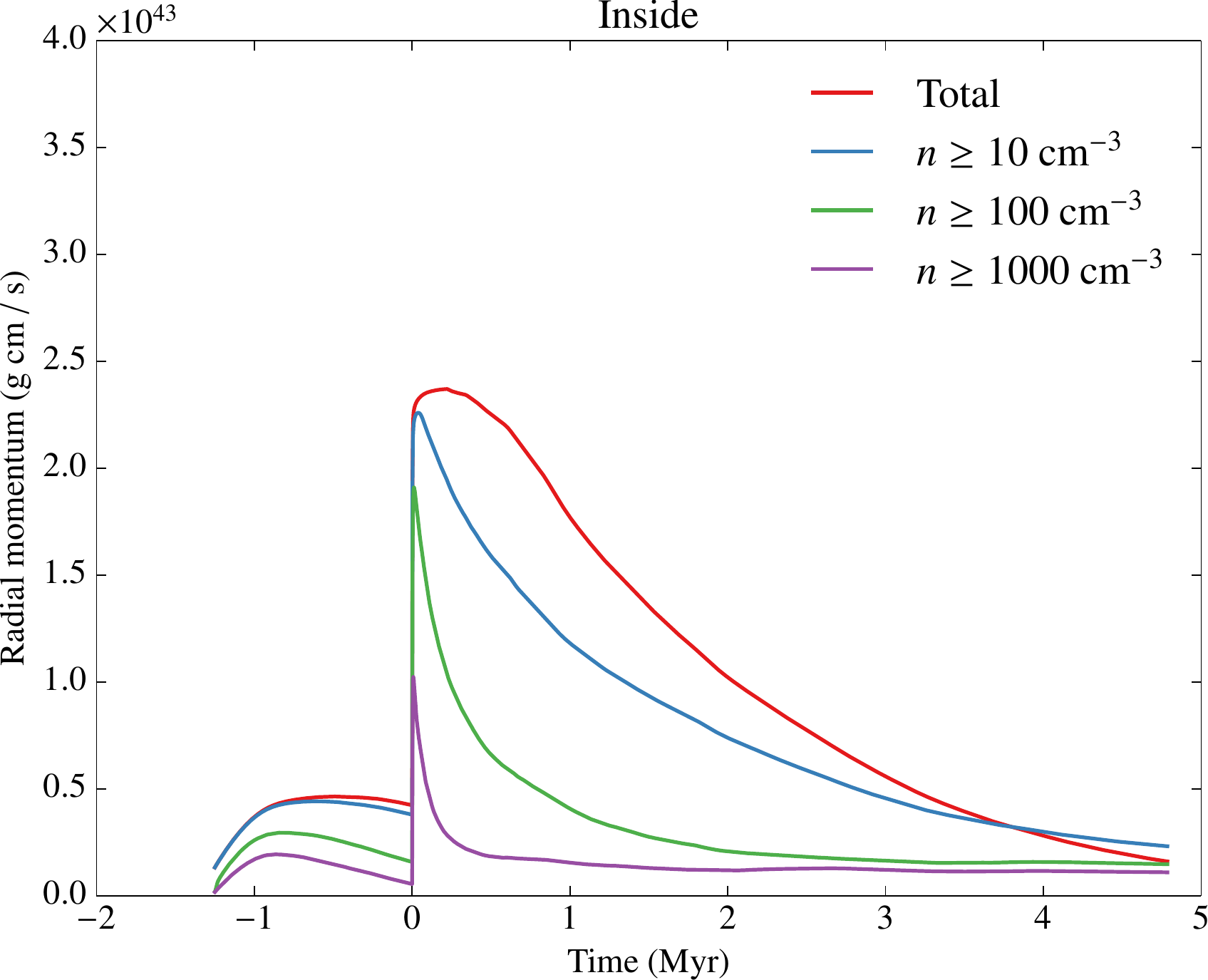}
               \end{center}
               \caption{MHD case. Evolution of momentum for densities above thresholds
                  $10, 100,$ and $1000\ \mathrm{cm}^{-3}$ in the outside, border,
                  and inside cases.}
               \label{graph-mom-thr-mhd}
            \end{figure}

            In addition to the value of the total momentum, it is important (as
            discussed above) to quantify the momentum distribution as a function
            of density.  Figure~\ref{graph-mom-thr-mhd} shows the momentum
            injected for the three density thresholds as a function of time for
            the MHD runs. These results should be compared with
            Fig.~\ref{graph-mom-thr}. There are more differences than for the
            total momentum, particularly in the inside run for which it is seen
            that the momentum delivered at high density just after the supernova
            explosion is about $20$ to $30\,\%$ higher than in the
            hydrodynamical case. This is in good agreement with the results
            obtained for the mass evolution (Fig.~\ref{graph-mass-mhd-thr}).

            Altogether, the various MHD runs presented in this section reveal
            that the magnetic field does not modify the total amount of momentum
            injected by supernovae into the ISM. It has a modest impact on the
            momentum that is injected into the dense gas, and therefore on the
            impact that supernovae may have in limiting star formation, if the
            supernova lies inside the dense gas when it explodes. The magnetic
            field has almost no impact if the supernova explodes outside the
            denser regions.

%         \subsubsection{Kinetic energy injection}
 
%            For the kinetic energy (figure \ref{graph-kinetic-turb-mhd}), the
%            evolution is still similar to our model, but messier because of the
%            presence of turbulence and magnetic field.
 
%            \begin{figure}
%               \begin{center}
%                  \includegraphics[width=80mm]{pic/turb/kin_mhd.pdf}
%               \end{center}
%               \caption{Kinetic energy injection: comparison between turbulent MHD
%               and our model. The vertical lines correspond to the first matter
%               outflow for each case (from left to right: border, outside, inside)}
%               \label{graph-kinetic-turb-mhd}
%            \end{figure}

\end{appendix}

\end{document}